\documentclass[twocolumn,showkeys,prd,nofootinbib,floatfix,preprintnumbers]{revtex4-1}

\usepackage[utf8]{inputenc}
\usepackage{multirow}
\usepackage{amsfonts,amsmath,amssymb} 
\usepackage{graphicx,graphics,color}
\usepackage{gensymb}
\usepackage{longtable}
\usepackage{bbding}
\usepackage{subfigure}
\usepackage[T1]{fontenc}
\usepackage{hyperref}
\usepackage[normalem]{ulem}
\hypersetup{
    colorlinks=true,
    linkcolor=blue}

  \def\aap{A\&A} 
  
  \def\prd{Phys. Rev. D}

\begin{document}

\title{Scalar Field Dark Matter Around Charged Black Holes}

\author{Yuri Ravanal}
\email{yuri.ravanal@usach.cl}
\affiliation{Departamento de F\'isica, Universidad de Santiago de Chile,\\Avenida V\'ictor Jara 3493, Estaci\'on Central, 9170124 Santiago, Chile}

\author{Gabriel G\'omez}
\email{gabriel.gomez.d@usach.cl}
\affiliation{Departamento de F\'isica, Universidad de Santiago de Chile,\\Avenida V\'ictor Jara 3493, Estaci\'on Central, 9170124 Santiago, Chile}

\author{Norman Cruz}
\email{norman.cruz@usach.cl}
\affiliation{Departamento de F\'isica, Universidad de Santiago de Chile,\\Avenida V\'ictor Jara 3493, Estaci\'on Central, 9170124 Santiago, Chile}
\affiliation{Center for Interdisciplinary Research in Astrophysics and Space Exploration (CIRAS), Universidad de Santiago de Chile, Avenida Libertador Bernardo O’Higgins 3363, Estación Central, Chile}

\begin{abstract}
In this paper, we investigate the behavior of a massive scalar field dark matter scenarios in the large mass limit around a central Reissner-Nordstr\"{o}m black hole. This study is motivated by observations from the Event Horizon Telescope collaboration, which does not exclude the possibility of the existence of such black holes. 
Through these inquiries, we uncover that the electric charge may significantly impact the scalar field profile and the density profile in the vecinty of the black hole. For the maximum electric charge allowed by the constraints of the Event Horizon Telescope, the maximum accretion rate decreases by $\thicksim$ 50 \% compared to the Schwarszchild case for marginally bound orbits. The maximum accretion rate of the massive scalar field is approximately $\dot M_{\text{SFDM}} \thicksim 10^{-8} M_{\odot} \;\text{yr}^{-1}$, which is significantly lower than the typical baryonic accretion rate commonly found in the literature. This implies that the scalar cloud located at the center of galaxies may have survived untill present times.

\end{abstract}

\maketitle

\section{Introduction}

There is robust evidence suggesting the presence of an invisible component in the observable Universe, known as dark matter (DM) \cite{WMAP:2003elm,Planck:2015fie,Planck:2018vyg,SDSS:2003tbn,BOSS:2013rlg,BOSS:2016wmc}. The most accepted and successful model to date is cold dark matter (CDM), which describes DM as a non-relativistic perfect fluid. However, CDM faces challenges when trying to explain observations at galactic and subgalactic scales. For instance, $N$-body simulations for CDM indicate a pronounced increase in the density profile (spike) as we approach the centers of galaxies. This is referred to as the universal or Navarro-Frenk-White (NFW) profile $\rho_{\text{NFW}}$ \cite{Navarro:1995iw}. In contrast, observations suggest constant density profiles forming a core toward the center \cite{Moore:1994yx,deBlok:2009sp}. This aspect is more evident in dwarf spheroidal and low-mass spiral galaxies \cite{Oh:2010mc}, as is the well-known discrepancy of 2-3 orders of magnitude in the number of satellite galaxies between observations and theoretical predictions \cite{Moore:1999nt,Bullock:2000wn}, or the famous problem known as ``too big to fail'' \cite{Boylan-Kolchin:2011qkt,Garrison-Kimmel:2014vqa}. However, some research suggests that taking into account the effects of baryonic feedback \cite{DelPopolo:2016emo,Dutton:2018nop,Dutton:2020vne} or even supernova feedback \cite{Sommer-Larsen:1998ppj} could help alleviate these tensions. Given all this, the idea of exploring new concepts arises. Some researchers propose modifying gravity in the low-acceleration regime \cite{Milgrom:1983ca}, while others suggest an alternative approach by changing the DM model, i.e., going beyond the pressureless CDM paradigm. In this work, we focus on the latter proposal.


Scalar field dark matter (SFDM) is a well-grounded candidate that arises as an extension of the standard model of particle physics and proposes DM is composed of ultralight bosonic particles with zero spin, spanning a mass range from 10$^{-22}$ eV to 1 eV. This model has gained prominence recently due to its potential to address some of these issues \cite{Hu:2000ke,Hui:2016ltb,Ferreira:2020fam}, as it possesses de Broglie wavelength on the order $\thicksim$ kpc. Within the SFDM model, there are various subcategories, including examples such as fuzzy dark matter (FDM) \cite{Hu:2000ke,Hui:2016ltb,Sin:1992bg,Ji:1994xh,Seidel:1990jh,Matos:1998vk,Hui:2019aqm,Dave:2023wjq,Pantig:2022sjb}, where the particle is typically associated with a mass on the order $\thicksim$ 10$^{-22}$ eV, self-interacting scalar field dark matter (SIDM) \cite{Goodman:2000tg,Peebles:2000yy,Arbey:2003sj,Boehmer:2007um,Lee:2008jp,Harko:2011xw,Rindler-Daller:2011afd,Suarez:2013iw,Urena-Lopez:2019kud,Lee:1995af}, involving coupling between particles, the well-known Axions \cite{Peccei:1977hh,Wilczek:1977pj,Weinberg:1977ma,Khlopov:1985fch,Bhattacharyya:2023kbh}, originally proposed to address the CP symmetry violation problem, and also Axion-Like-Particles (ALPs) proposed in other theories \cite{Marsh:2015xka}. The main distinction among the aforementioned models lies in the mass range or the associated coupling of these particles \cite{Kobayashi:2017jcf,Abel:2017rtm,Brito:2017wnc,Brito:2017zvb,Dave:2023egr}.

A fascinating aspect of SFDM models is that the stationary solutions of these classical bosonic fields, known as solitons, could be present in galactic centers, providing a plausible explanation for the observed constant density profiles \cite{Arbey:2001qi,Schive:2014hza,Marsh:2015wka,Schwabe:2016rze,Mocz:2017wlg,Levkov:2018kau,2011PhRvD..84d3531C,2011PhRvD..84d3532C}. These solitonic cores would form structures ranging from 1 kpc to 20 kpc. In the case of FDM, it would be attributed to quantum pressure, arising from the uncertainty principle, while for SIDM, it would be due to self-interaction (repulsive). However, these would not be the only mechanisms for the formation of gravitationally bound structures at the galactic level. In fact, galactic halos are expected to be in equilibrium due to virialized motion \cite{2015eaci.book.....S}. Additionally, these models could also explain the observed vortices in galaxies \cite{Rindler-Daller:2011afd,Peebles:1969jm,Kain:2010rb}. Although current measurements \cite{Kobayashi:2017jcf,Irsic:2017yje,Rogers:2020ltq,Bar:2021kti} suggest a minimum mass limit of around $ m \gtrsim 10^{-22}$ eV for the FDM model.

The vast majority of galaxies in the Universe host supermassive black holes (SMBHs) at their centers \cite{Kormendy:1995er,Ferrarese:2004qr,Narayan:2005ie}. The prevailing notion is that these black holes (BHs) are surrounded by a solitonic core, which, in turn, is enveloped by a halo of DM. Some researches have suggested the possibility that these BHs possess a non-zero value of net electric charge \cite{Zakharov:2014lqa,Juraeva:2021gwb}. Along the same lines, the Event Horizon Telescope (EHT) collaboration has recently provided constraints on the mass to charge ratio $q = |Q/M|$ \cite{EventHorizonTelescope:2021dqv,EventHorizonTelescope:2022xqj}. Initially, they captured the shadow image of M87$^{\star}$ \cite{EventHorizonTelescope:2021dqv} and later did the same for Sagittarius A$^{\star}$ (SgrA$^{\star}$) \cite{EventHorizonTelescope:2022xqj,Vagnozzi:2022moj}. These studies indicate that the charge to mass ratio for M87$^{\star}$ falls within the range of $ q \in [0,0.9]$ \cite{EventHorizonTelescope:2021dqv}, while for SgrA$^{\star}$, it is $q \in [0,0.84]$ \cite{EventHorizonTelescope:2022xqj}. Nevertheless, it is widely disseminated in the community that BHs are electrically neutral since the surrounding plasma would rapidly discharge them. It is important to note that this article will not delve into the origin of electric charge or the timescales during which a BH might remain electrically charged. Therefore, we strictly adhere to the observational constraints from EHT.

The nature of DM remains unknown. However, with the first detection of gravitational waves (GWs) \cite{LIGOScientific:2016aoc}, another research frontier opened up in this field. It is possible that within the signal of GWs, relevant information about the medium in which BHs are immersed may be inferred, making it a feasible probing option to explore the nature of DM \cite{Eda:2013gg,Lacroix:2012nz,Kavanagh:2020cfn,Gomez:2016iod,Bar:2019pnz,Saurabh:2020zqg,Boudon:2023vzl,Chakrabarti:2022owq,Berezhiani:2023vlo,Rahman:2023sof}. Nevertheless, the impact that DM has on the GW signal depends on the its properties, such as the density profile. Therefore, it is of vital importance to carry out accurate and self-consistent modeling of DM around black holes. For instance, density profiles could differ depending on whether it is CDM \cite{Sadeghian:2013laa,Gondolo:1999ef} or SFDM \cite{Barranco:2011eyw,Clough:2019jpm,Bamber:2020bpu,Hui:2019aqm,Aguilar-Nieto:2022jio,Brax:2019npi,Vicente:2022ivh,Cardoso:2022nzc}, and to this day, the discussion about the existence of ``scalar hair'' continues \cite{Jacobson:1999vr,Hui:2019aqm,Brihaye:2021phs,Hong:2020miv}.

Currently, there exists an extensive literature on the study of scalar fields in Schwarzschild space-time (see, for example, \cite{Unruh:1976fm,Detweiler:1980uk,Hui:2019aqm}); however, only a small handful of studies delve into the realm of ``non-standard metrics'' (see e.g. \cite{Benone:2014qaa,Hong:2019mcj,Ovgun:2023wmc,Babichev:2008jb}). It is precisely within this context that our research is aimed.

In this study, we specifically focus on investigating the behavior of a non-interacting massive scalar field DM in the large-mass limit ($ m \gg 10^{-22}$ eV) around a Reissner-Nordstr\"{o}m (RN) BH \cite{reissner1916eigengravitation}. Throughout this work, we ignore the backreaction effect of the SF on the metric. Therefore, we adopt the test-fluid approximation for the sake of simplicity. We establish a direct connection between these results and our previous research \cite{Ravanal:2023ytp}, following a similar path to that taken by \cite{Brax:2019npi}. We uncover that the influence of the electric charge $Q$ becomes relevant in regions close to  marginally bound orbits\footnote{More precisaly, this zone extends from the marginally bound radius $r_{\text{mb}}$ to the radius of the innermost stable circular orbit $r_{\text{isco}}$, that is, $ r_{\text{mb}} \lesssim r \lesssim r_{\text{isco}}$.}. Considering this argument \cite{Ravanal:2023ytp,Beheshti:2015bak,Hod:2013mgr}, a maximum accretion rate of the scalar field DM of $\dot M_{\text{SFDM}} \thicksim 10^{-8} M_{\odot}\;\text{yr}^{-1}$ was obtained, with a decrease of approximately 50 \% when $Q$ becomes significant compared to the uncharged case. Within the constraints allowed by the EHT, the accretion in the non-interacting case, which corresponds to a free-falling, is higher compared to the SIDM case \cite{Feng:2021qkj,Ravanal:2023ytp}, since in the latter, the repulsive self-interaction slows down the DM fall. This feature determines the lifetime and the mass of the extended cloud and have critical consequences for the observed shadow radius \cite{Gomez:2024ack}. Furthermore, a scalar field profile $\phi \propto r^{-3/4}$ was observed, which coincides with the same power law as other studies in the same regime of interest \cite{Hui:2019aqm,Clough:2019jpm,Bucciotti:2023bvw}. Finally, a change in the power-law exponents $\rho \propto r^{-3/2}$ and $\rho \propto r^{-1}$ for the density profile in the non-interactive and interactive cases, respectively, was observed. This behavior has also been documented in previous studies \cite{Gondolo:1999ef,Hui:2019aqm,Brax:2019npi,DeLuca:2023laa}. Evidently, this change originates from the nature of the studied fields. These exponents fall within the usual range considered in various power-law models for dark matter density profiles \cite{Ferreira:2020fam}.


This paper is structured as follows. In section \ref{sec:II}, we describe the theoretical framework, including the DM model, conditions of steady-state accretion, and the resulting scalar field profile. In section \ref{sec:III}, we establish a direct connection with our previous work \cite{Ravanal:2023ytp}, and compare our main findings with other works. Finally, in section \ref{sec:IV}, we provide a detailed discussion of our results and their potential implications.

\section{Dark Matter Scalar Field}
\label{sec:II}
The relativistic action of a real scalar-field minimally coupled to gravity is given by
\begin{equation}
    S= S_{\text{EH}}+S_\phi = \int d^4x \sqrt{-g} \left[ \frac{\mathcal{R}}{16 \pi} - \frac{1}{2} g^{\mu\nu} \partial_\mu\phi \partial_\nu\phi
- V(\phi) \right].\label{action relativi}
\end{equation}
Here $S_{\text{EH}}$ represents the Einstein-Hilbert action, $S_\phi$ corresponds to the action for a single real SF $\phi$, $g_{\mu \nu}$ represents the metric, $\mathcal{R}$ denotes the Ricci scalar, $g$ stands for the determinant of the metric $g_{\mu \nu}$, and $V(\phi)$ is the SF potential. In this study, we adhere to metric signature conventions of $(-,+,+,+)$ and use units where $4\pi\epsilon_{0} = G = c = \hbar = 1$.

The equation of motion for the SF, derived from Eq. (\ref{action relativi}), takes the following form
\begin{equation}
\frac{\delta S_\phi }{\delta \phi} = 0 \;\;\; \Longrightarrow \;\;\; \Box \phi - \frac{dV}{d\phi} = 0,\label{Equation motion}
\end{equation}
where the covariant d’Alembertian $\Box$ is defined as $\Box = \nabla^{\mu}\nabla_{\mu} = g^{\mu\nu}\nabla_{\mu}\nabla_{\nu}$. For the potential $V(\phi) = m^2 \phi^2/2$, we derive the well-known Klein-Gordon equation $(\Box - m^2)\phi = 0$. Additionally, starting from Eq. (\ref{action relativi}), we can calculate the energy-momentum tensor for the SF, expressed as $T^{\mu \nu} = (2/\sqrt{-g}) \; \delta S_\phi / \delta g_{\mu \nu}$.

In this article, we consider the large-mass limit \cite{Hui:2019aqm,Brax:2019fzb} given by
\begin{equation}
    m \gg 6.7 \times 10^{-19} \; \left( \frac{M}{10^8 M_\odot} \right)^{-1} {\rm eV},
    \label{eq. limite masa grande}
\end{equation}
where the characteristic length scale of the system surpasses the Compton wavelength $\lambda_C \thicksim 1/m \ll r_{\text{h}}$. Here $r_{\text{h}} $ is the radius of the event horizon. In this regime, we can neglect the quantum pressure $\Phi_{Q}$. This pressure stems from Heisenberg’s uncertainty principle $\Delta x \Delta p \thicksim 1$ and is insignificant at galactic and subgalactic scales within our regime. Consequently, the SFDM cloud is governed by self-gravity at these scales, reaching a virial equilibrium state where the kinetic energy of the particles equals the gravitational potential energy.

Before deriving the master equations, we outline the primary physical assumptions made in this work, aiming to enhance clarity:
\begin{itemize}
    \item Radial accretion flows on static spherically symmetric BHs.
    \item large scalar mass limit where the Compton wavelength $1/m$ is smaller than the BH size.
    \item Test-fluid approximation where the backreaction of the scalar cloud to the spacetime metric is neglected. 
\end{itemize}

\subsection{Spherically
symmetric space-times}
\label{Symetric space-times}

For a spherically symmetric spacetime the metric takes the form
 \begin{equation}
 ds^2 = - f(r) dt^2 + g(r) dr^2 + r^2  d\vec\Omega^2,\label{metric SS}
 \end{equation}
 where the metric functions $f(r)$ and $g(r)$ for a RN metric are given by
 \begin{equation}
 f(r)=\frac{1}{g(r)}=1-\frac{2M}{r}+\frac{Q^2}{r^2},\label{coef}
 \end{equation}
with the corresponding horizons
 \begin{equation}
 r_{\pm} = M \pm \sqrt{M^2-Q^2}.\label{horizont}
 \end{equation}
$M$ represents the mass of the BH, and $Q$ denotes the electric charge. The BH metric exhibits two event horizons: $r_+$ corresponds to the outer horizon (of primary interest in this study), while $r_-$ corresponds to the Cauchy horizon. The Schwarzschild solution is readily obtained when $Q = 0$, and the scenario $Q=M$ represents the extremal case. The metric functions can be easily expressed as a function of dimensionless quantities through the following variable transformation
\begin{equation}
x=\frac{r}{M} \geq 1,\; \text{and} \;\;\;q=\frac{Q}{M},\label{Change Variables}
\end{equation}
where $x$ denotes the new radial coordinate, and $q$ represents the charge-to-mass ratio. Utilizing these variables, we can describe the horizons in the following manner
\begin{equation}
    x_{\pm} = 1 \pm \sqrt{1-q^2} : \;\;\; 0\leq q \leq 1.\label{new horizont}
\end{equation}
It is important to note that for the sake of completeness, there are two other possible cases in Eq. (\ref{new horizont}). The first one is when $q > 1$, which describes a naked singularity\footnote{It is important to mention that even if it were a naked singularity, it could exhibit a photon sphere and therefore be an indication of a shadow \cite{Vagnozzi:2022moj}.} without an event horizon, and the second one is when $q^2$ is negative. However, in this work, we will not consider these two cases, as we will focus solely on the constraints imposed by the EHT \cite{EventHorizonTelescope:2021dqv,EventHorizonTelescope:2021srq,EventHorizonTelescope:2022xqj,Vagnozzi:2022moj}.

In this scenario, we can divide the physical regions of interest around the BH into three zones as follows:
\begin{itemize}
    \item {\bf region near the BH} ($r_{+} < r < r_{\text{NL}}$): This region extends from the horizon $r_+$ to a non-linear radius $r_{\text{NL}}$. It is denoted as the strong-gravity regime, where the metric functions in the vicinity of the BH are defined by the Eq. (\ref{coef}).
    \item {\bf intermediate region} ($r_{\text{NL}} \ll r \ll r_{\text{sg}}$): This region lies between the non-linear radius $r_{\text{NL}}$ and the transition radius $r_{\text{sg}}$. The latter is defined as the radius at which the self-gravity of the DM scalar cloud equals the BH gravitational potential. This radius is very important because it marks the position at which the scalar field profile is affected by the BH gravity. For realistic DM models, we expect this radius to be greater than the BH horizon. Therefore, in this region, we are in the weak field regime ($r_{\text{NL}} \ll r$), where the line element takes the form of $ds^2 = -(1+2\Phi)dt^2 + (1-2\Phi)dr^2 + r^2  d\vec\Omega^2$ with $\Phi \ll 1$. In this regime, the gravitational potential of the BH is the usual $\Phi = -M/r$.
    \item {\bf region far from the BH} ($r\gg r_{\text{sg}} $): This region lies beyond the transition radius $r_{\text{sg}}$. Therefore, we are in the weak-field regime, where the line element takes the aforementioned form, but the contributions to the metric potential are mostly due to the self-gravity of the DM cloud. Consequently, the gravitational potential $\Phi$ is governed by the scalar field's Poisson equation 
    \begin{equation}
        \nabla^2\Phi = 4\pi\rho_{\phi},
        \label{eq. poisson sf}
    \end{equation}
    where $\rho_\phi$ stands for the energy density of the SF.
    It is important to note that throughout this manuscript, our region of interest mainly lies in the vicinity and intermediate region of the BH. The region far from the BH plays a role in section \ref{self interactive scalar} where the solution must match the well-known hydrostatic equilibrium configuration.
\end{itemize}
\subsection{Free Scalar Field}
\label{Free scalar field}
We star by examining the profile of the free SF flow around the RN-BH.

\subsubsection{Equations of motion}
\label{equation motion}
The relativistic action of the SF (\ref{action relativi}) is expressed in terms of the metric (\ref{metric SS})
\begin{multline}
S_\phi = \int dt dr d\theta d\varphi \sqrt{f g} r^2 \sin\theta \left[ \frac{1}{2 f}
\left( \frac{\partial \phi}{\partial t} \right)^2 - \frac{1}{2 g}
\left( \frac{\partial \phi}{\partial r} \right)^2 \right.\\
\left. - \frac{1}{2  r^2} \left( \frac{\partial \phi}{\partial \theta} \right)^2
- \frac{1}{2  r^2 \sin^2\theta} \left( \frac{\partial \phi}{\partial \varphi} \right)^2 - V(\phi) \right].
\label{eq.action SF real SS}
\end{multline}

It is useful to express the real SF $\phi$ in terms of a complex SF $\psi$
\cite{Salehian:2021khb,Brax:2019fzb}
\begin{equation}
    \phi = \frac{1}{\sqrt{2m}} \left( e^{-imt} \psi + e^{imt} \psi^*\right),
    \label{eq. SF real }
\end{equation}
where $\psi^*$ represents the complex conjugate of $\psi$. This decomposition is particularly useful in the non-relativistic regime because it allows us to separate the fast oscillations dominated by the frequency from the slower dynamics described by the field 
$\psi$. Therefore, $\psi$ is a slowly varying function of time and space compared to the dominant frequency of the system $\omega\sim m$. Following the same idea expressed in the previous paragraph, in the large mass limit, we have $\dot\psi \ll m \psi$ (slow evolution condition), where the overdot denotes the time derivative. On the other hand, we can observe that $\psi$ exhibits a global symmetry of the $U(1)$ group, signifying its invariance under continuous phase rotations $\xi$. A relevant observation about the real SF $\phi$ is that there is no conserved (Noether) charge, in contrast to a complex SF $\psi$, by virtue of Noether's theorem\footnote{In the non-relativistic regime, one can see that the conserved current associated with such a symmetry is proportional to $m \psi \psi^{*}$, which corresponds to the conservation of matter density (see e.g., \cite{Brax:2019fzb}).}
It is important to note that the global $U(1)$ symmetry for $\psi$ is the result of having $\phi$ invariant under the transformation \cite{Brax:2019fzb}
\begin{equation}
\left.
\begin{array}{c}
\psi \rightarrow \psi e^{i \xi} \:\: , t \rightarrow t + \xi /m \\
\psi^* \rightarrow \psi^* e^{-i \xi} \:\: , t \rightarrow t + \xi /m
\end{array}
\right\}
\vspace{1cm}
\Longrightarrow \phi \longrightarrow \phi.
\label{eq. invarianza U(1)}
\end{equation}

Taking into account all mentioned above, we can bring our problem into the large mass limit. We can express the action of the complex SF $\psi$ as follows
\begin{multline}
S_\psi = \int dt dr d\theta d\varphi \sqrt{f g} r^2 \sin\theta \left[ \frac{1}{2 f}
( i \dot\psi \psi^* - i \psi \dot\psi^* \right. \\ + m \psi \psi^* ) - \frac{1}{2 m g}
\frac{\partial \psi}{\partial r} \frac{\partial \psi^*}{\partial r}
- \\ \frac{1}{2 m r^2} \frac{\partial \psi}{\partial \theta}
\frac{\partial \psi^*}{\partial \theta} \left. - \frac{1}{2 m r^2 \sin^2\theta}
\frac{\partial \psi}{\partial \varphi} \frac{\partial \psi^*}{\partial \varphi}
- \frac{m}{2} \psi \psi^* \right].
\label{eq.action SF psi SS}
\end{multline}
Usually, rapid oscillations $e^{\pm 2imt}$ are discarded, as their average is approximately zero under the previously described assumptions. Finally, we can derive the Euler-Lagrange equations for the field $\psi$
\begin{equation}
    i \dot\psi = - \frac{f}{2m} \left [ \frac{1}{\sqrt{fg}} \vec\nabla_r \cdot \left (\sqrt\frac{f}{g} \vec\nabla_r \psi \right)+\nabla^2_{\theta,\varphi}\psi \right]
+ m \frac{f-1}{2} \psi.
\label{eq. E-L psi}
\end{equation}
Here, $\vec\nabla_r$ and $\vec\nabla_{\theta,\varphi}$ represent the radial and angular components of the nabla operator, respectively. On the other hand, we can recover the non-relativistic version for $\psi$ at large distances, as the metric functions are recovered in the weak-field limit (intermediate region in section \ref{Symetric space-times})
\begin{equation}
    r \gg r_{\text{h}} \; \; : \; i \dot\psi = - \frac{\vec\nabla^2 \psi}{2m} + m \Phi \psi.
    \label{eq. E-L psi N-Relati}
\end{equation}
We can observe that the above expression is a Schrödinger-type equation. It is possible to approach this problem within the fluid description using the Mandelung transformations \cite{madelung1927quantum}
\begin{equation}
    \psi = \sqrt{\frac{\rho}{m}} e^{i s} \; \; \;, \; \; \; \rho = m| \psi |^2,
\label{eq. transformacion maledung}
\end{equation}
where $\sqrt{\rho/m}$ represents the amplitude and $s$ is a phase, $\rho$ serves as the SF matter density, and the velocity field is defined as follows
\begin{equation}
    \vec{v} = \frac{\vec\nabla s}{m},
    \label{eq. definicion velocidad}
\end{equation}
in this description, the flow is defined as irrotational. Furthermore, with the previously mentioned definitions, we can express the real scalar field profile as follows\footnote{It is important to mention that many authors use complex scalar fields instead. Nevertheless,  adding a complex conjugate in such solutions, one can obtain the corresponding real field and \textit{vice versa}. For instance, the density profiles between both fields differ by a factor of $1/2$. Then, our setup can safely cover complex fields.}
\begin{equation}
    \phi = \frac{\sqrt{2\rho}}{m} \cos(m t - s).
    \label{eq. SF real Maledung}
\end{equation}
It is possible to express the action $S_{\psi}$ in terms of the new quantities $\rho$ and $s$ through the Madelung transformations. The new action now reads:
\begin{widetext}
\begin{align}
S_{\rho,s} = \int dt dr d\theta d\varphi \sqrt{fg} r^2 \sin\theta \biggl \lbrace
- \frac{\rho \dot s}{m f} - \frac{1}{2 m^2 g} \left[ \frac{1}{4\rho} \left( \frac{\partial \rho}{\partial r} \right)^2
\!\! + \rho \left( \frac{\partial s}{\partial r} \right)^2 \right]
- \frac{1}{2 m^2 r^2} \left[ \frac{1}{4\rho} \left( \frac{\partial \rho}{\partial \theta} \right)^2
\right. \left. + \rho \left( \frac{\partial s}{\partial \theta} \right)^2 \right] \nonumber\\
- \frac{1}{2 m^2 r^2 \sin^2\theta} \left[ \frac{1}{4\rho}
\left( \frac{\partial \rho}{\partial \varphi} \right)^2 \!\! + \rho \left( \frac{\partial s}{\partial \varphi} \right)^2 \right] + \frac{\rho}{2f} - \frac{\rho}{2} \biggl \rbrace.
\label{eq.action rho y s}
\end{align}
\end{widetext}
As mentioned earlier, in the large-mass limit, neglecting spatial gradients $|\vec \nabla \rho | \ll m \rho$ is equivalent. However, since the phase $s$ is of the order of $m$, the action is consequently reduced to the following form
\begin{multline}
S_{\rho,s} = \int dt dr d\theta d\varphi \sqrt{f g} r^2 \sin\theta \Biggl[ - \frac{\rho \dot s}{m f} - \frac{\rho}{2 m^2 g} \left( \frac{\partial s}{\partial r} \right)^{\!\! 2} \\ - \frac{\rho}{2 m^2 r^2} \left( \frac{\partial s}{\partial \theta} \right)^{\!2}
- \frac{\rho}{2 m^2 r^2 \sin^2\theta}\left( \frac{\partial s}{\partial \varphi} \right)^{\!\! 2} \\ + \frac{\rho (1-f)}{2f} \Biggr].
\label{eq.action rho y s limite-m}
\end{multline}
We can obtain the equations of motion from action (\ref{eq.action rho y s limite-m}), that is, $\delta S/\delta s = 0$, resulting in 
\begin{equation}
    \dot\rho + f\left [ \frac{1}{\sqrt{fg}} \vec\nabla_r \cdot \left( \sqrt{\frac{f}{g}} \rho \frac{\vec\nabla_r s}{m} \right)+\vec\nabla_{\theta,\varphi}\cdot\left(\frac{\rho}{m}\vec\nabla_{\theta,\varphi} s\right) \right ] = 0.
    \label{eq. continuidad 1}
\end{equation}
The same applies to $\delta S/\delta \rho = 0$:
\begin{equation}
    \frac{\dot s}{m} + \frac{f}{2}\left[ \frac{1}{g}\frac{(\vec\nabla_r s)^2}{m^2}+\frac{(\nabla_{\theta,\varphi}s)^2}{m^2} \right]= \frac{1-f}{2}.
    \label{eq. euler}
\end{equation}
By taking the gradient of Eq. (\ref{eq. euler}) and using the definition of the velocity field (\ref{eq. definicion velocidad}), in Eqs. (\ref{eq. continuidad 1}) and (\ref{eq. euler}) we have
\begin{align}
\dot{\rho} + f \left [ \frac{1}{\sqrt{fg}} \vec\nabla_r \cdot \left( \sqrt{\frac{f}{g}} \rho v_r \right)+\vec\nabla_{\theta,\varphi}\cdot\left(\rho v_{\theta,\varphi}\right) \right ] = 0 ,
\label{eq.continuidad 2}
\end{align}
\begin{align}
\dot{\vec v} + \vec\nabla \left [ \frac{f}{2}\left( \frac{v^2_r}{g}+v^2_{\theta,\varphi} \right) \right ] =
- \frac{\vec\nabla f}{2} .
\label{eq. euler 2}
\end{align}
For very large distances ($r \gg r_{\text{h}}$), the metric functions are recovered in the weak-field limit, and by using vector identities, we obtain
\begin{align}
\dot\rho + \vec\nabla \cdot ( \rho \vec{v} ) &= 0 ,
\label{eq.coninuidad 3}
\end{align}
\begin{align}
\dot{\vec v} + (\vec v \cdot \vec\nabla) \vec v &= - \vec\nabla \Phi .
\label{eq.euler3}
\end{align}
Eqs. (\ref{eq.coninuidad 3}) and (\ref{eq.euler3}) represent the continuity and Euler equations, respectively. Both equations correspond to the classical limit that governs fluid dynamics. It is important to mention that Euler's equation (\ref{eq.euler3}), which lacks a pressure term, corresponds to the free motion of particles under the influence of a gravitational potential $\Phi$. It is important to mention that, starting from Eq (\ref{eq. E-L psi N-Relati}) and using the Madelung transformations Eqs. (\ref{eq. transformacion maledung}) and (\ref{eq. definicion velocidad}), when separating the real and imaginary parts, we obtain the same equations (continuity and Euler) in the non-relativistic limit. We neglect the quantum pressure $\Phi_{Q} = - (\nabla^2 \sqrt{\rho})/(2m^2\sqrt{\rho})$, as we are in the large-mass limit, where the momentum are considerably smaller compared to the rest mass. 

\subsubsection{Steady state}
\label{staedy state}
We can find stationary solutions to Eqs. (\ref{eq.continuidad 2}) and (\ref{eq. euler 2}), and restricting to spherical symmetry (\ref{metric SS}), we obtain the following solution to the continuity equation (\ref{eq.continuidad 2})
\begin{equation}
    \sqrt\frac{f}{g} r^2 \rho v_r = F ,
\label{eq.flujo}
\end{equation}
where $F<0$ is defined as an inward flux per unit solid angle; in other words, DM falls into the BH steadily.

From the Euler equation (\ref{eq. euler 2}), we obtain
\begin{equation}
    v_r = \pm \sqrt{\frac{g (1-f)}{f}}\;\;\;,\;\;\; mv_r = \frac{\partial s}{\partial r}.
\label{eq.vr schw}
\end{equation}
In our case, we choose the negative sign solution, as DM particles fall radially towards the BH. Near the horizon, we observe a divergence when $f \to 0 $, which is caused by the use of non-regular coordinates. Furthermore, we can see that as the radius tends to infinity, the radial velocity tends to zero, which sets the boundary condition of the problem.

We can obtain the matter density of the SF from Eqs. (\ref{eq.flujo}) and (\ref{eq.vr schw}). This expression is valid in the large-mass limit, where we observe that massive particles fall radially toward the BH, regardless of their mass. They start at rest from infinity and free-fall into the vicinity of the BH
\begin{equation}
    \rho = - \frac{F}{r^2 \sqrt{1-f}} .
\label{eq.rho schw}
\end{equation}

When we approach the BH horizon, we notice that $f \to 0$. At this point, we observe that $\rho$ is finite and equals $ -F/r_{\text{h}}^2$. In contrast, when we are at very large distances, $f$ approaches unity, and consequently, $\rho \to 0$ because the quadratic term decreases faster than the metric function $f$.

\subsubsection{Scalar field profile}
\label{Scalar field profile}
\begin{figure}
    \centering
    \includegraphics[scale=0.335]{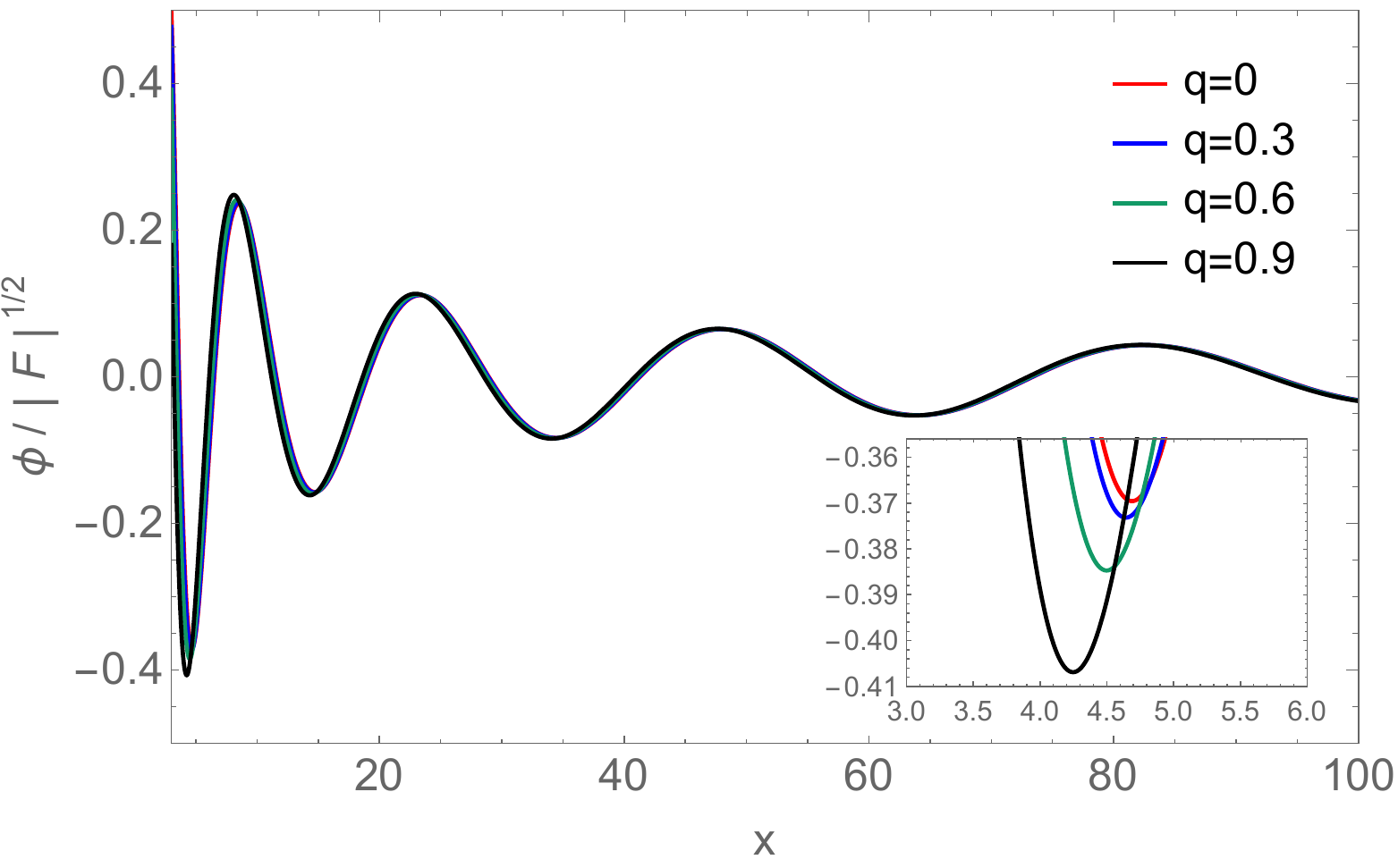}
    \caption{Normalized scalar field profile $\phi (x) / |F|^{1/2}$ as a function of the radial coordinate $x$, for different values of the charge-mass ratio $q$, as given by Eq. (\ref{eq.profile SF}). The zoomed-in window pertains to the region between the photon sphere $x_{\text{ph}}$ and the innermost stable circular orbit $x_{\text{isco}}$, which uncovers the effect of the BH charge on the scalar profile.}
    \label{SF profile}
\end{figure}

\begin{figure}
    \centering
    \includegraphics[scale=0.34]{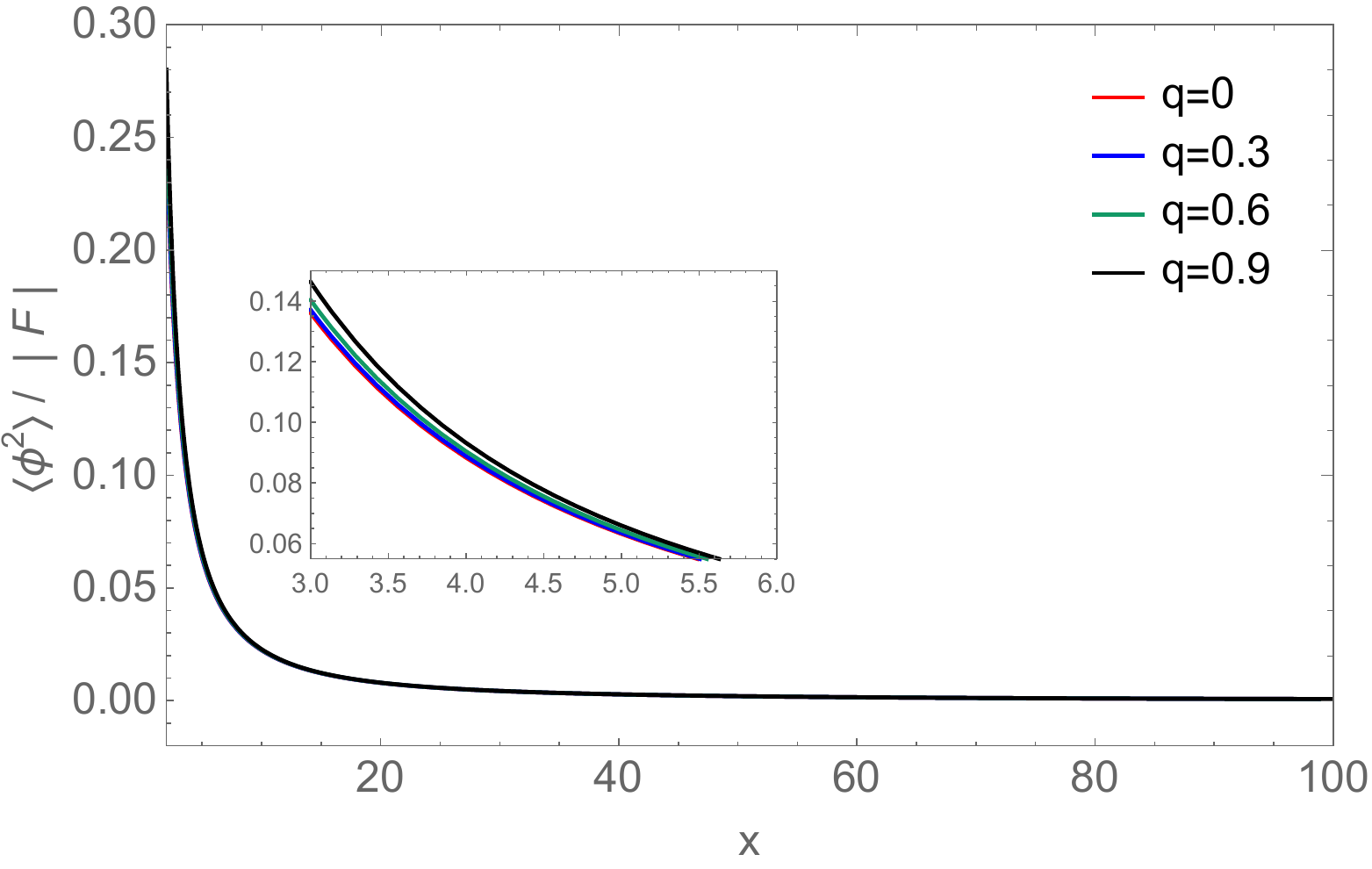}
    \caption{Normalized time-averaging of the square scalar field profile $ \langle \phi^2 \rangle / |F|$ as a function of the radial coordinate $x$, for different values of the charge-mass ratio $q$, as derived from Eq. (\ref{eq.profile SF}). The zoomed-in window pertains to the region between the photon sphere $x_{\text{ph}}$ and the innermost stable circular orbit $x_{\text{isco}}$.}
    \label{SF promedio}
\end{figure}

It is possible to rewrite the SF profile, considering the stationary solutions imposed in the previous section. This is achieved by substituting Eq. (\ref{eq.rho schw}) into Eq. (\ref{eq. SF real Maledung}), and the phase $s$ is obtained from the expressions in (\ref{eq.vr schw}). This gives
\begin{equation}
    \phi = \frac{1}{2mM}\sqrt{\frac{-8F}{x^2\sqrt{1-f}}}\cos(mt-s).
    \label{eq.profile SF}
\end{equation}
This  expression represents a harmonic ingoing wave propagating towards the BH, as shown in Fig. (\ref{SF profile}). The SF profile exhibits then oscillatory behavior\footnote{This feature is found for both real and complex SFs: in the latter case, we observe this behavior in the real part of the SF profile. It is important to note that the oscillatory behavior is affected by its mass (or equivalently its frequency), and in the limit of light masses, the oscillation occurs at slightly larger scales compared to heavier masses \cite{Hui:2019aqm,Clough:2019jpm,Bucciotti:2023bvw}.}. As we approach the BH, its amplitude increases or in the opposite case, it decreases, pimarily due to the denominator within the square root. We can observe that the effects of $q$ become significant in the vicinity of the BH; indeed, as $q$ increases, an increase in amplitude is observed, mainly because a BH with significant charge has a smaller event horizon compared to the uncharged case. To further clarify this idea, it is convenient to remove the temporal dependence of the SF profile\footnote{ When a field is oscillatory, the usual approach is to average it over the oscillation cycles, i.e., $\langle \; \cos^2 \; \rangle \to 1/2$.} (\ref{eq.profile SF}), as shown in Fig. (\ref{SF promedio}). These curves essentially represent the amplitude of the SF profile. Once more, it is observed that as the charge $q$ increases, the amplitude near the black hole also increases. At this point, we bring up Jacobson's results \cite{Jacobson:1999vr}, who obtained a non-vanishing solution for a massless scalar field by imposing the boundary condition of a non-zero time derivative far away from the BH. He obtained SF profile that behaves as $\phi \propto r^{-1}$ for large radii. In our case of large-mass limit framed on DM scenarios, the decay behaves differently, specifically as $\phi \propto x^{-3/4}$ for large radii.


\subsubsection{Density profile}
\label{density profile}
From the relativistic action (\ref{action relativi}), it is possible to obtain the energy-momentum tensor $T_{\alpha \beta} = \partial_{\alpha}\phi \partial_{\beta} \phi + g_{\alpha \beta}\mathcal{L}_{\phi}$, where $\mathcal{L}_{\phi}$ is the lagrangian of the SF. Under the symmetry of our problem (\ref{metric SS}), we can express the energy density of the SF $\rho_\phi$, which is associated with the time-time component of the energy-momentum tensor
\begin{equation}
    \rho_\phi \equiv - T^t_t = \frac{1}{2f} \left( \frac{\partial\phi}{\partial t} \right)^2
+ \frac{1}{2g} \left( \frac{\partial\phi}{\partial r} \right)^2 + V(\phi).\label{rho_scalar}
\end{equation}
Replacing the SF profile (\ref{eq. SF real Maledung}) in the previous equation and considering the large mass limit, we obtain
\begin{equation}
    \rho_{\phi} = \rho \left[ \frac{(2-f)}{f} \sin^2(mt-s) + \cos^2(mt-s)\right].
    \label{eq:rho SF libre}
\end{equation}
Finally, expressing it in terms of the flux $F$ and averaging over the rapid oscillations with a period of $2\pi/m$, we obtain:
\begin{equation}
    \langle \rho_{\phi} \rangle = -\frac{4F}{(2M)^2 x^2 f \sqrt{1-f}}.
    \label{eq. rho promedio SF libre}
\end{equation}
It is important to note that $q$ is constrained between 0 and 0.9 according to the EHT \cite{EventHorizonTelescope:2021dqv,EventHorizonTelescope:2021srq,EventHorizonTelescope:2022xqj,Vagnozzi:2022moj}. With all of this in mind, we can study the behavior of the density profile of this massive SF around a RN-BH. The average energy density $\langle \rho_{\phi} \rangle$ diverges as we approach the horizon. This divergence is caused by the term $1/f$, as $f \to 0$, as a result of using non-regular coordinates\footnote{We can easily manage this using Eddington coordinates instead. However, for astrophysical purposes, we keep the use of Schwarzschild coordinates throughout this paper.}. At large distances, we have $\langle \rho_{\phi} \rangle \propto x^{-3/2}$ and $|\;v\;| \propto x^{-1/2}$. This can be interpreted as a free particle falling from infinity with a velocity $\propto r^{-1/2}$ due to the conservation of energy, similar to virialized DM halos $v_r^2 \thicksim \Phi \thicksim M/r$.

\begin{figure}
    \centering
    \includegraphics[scale=0.33]{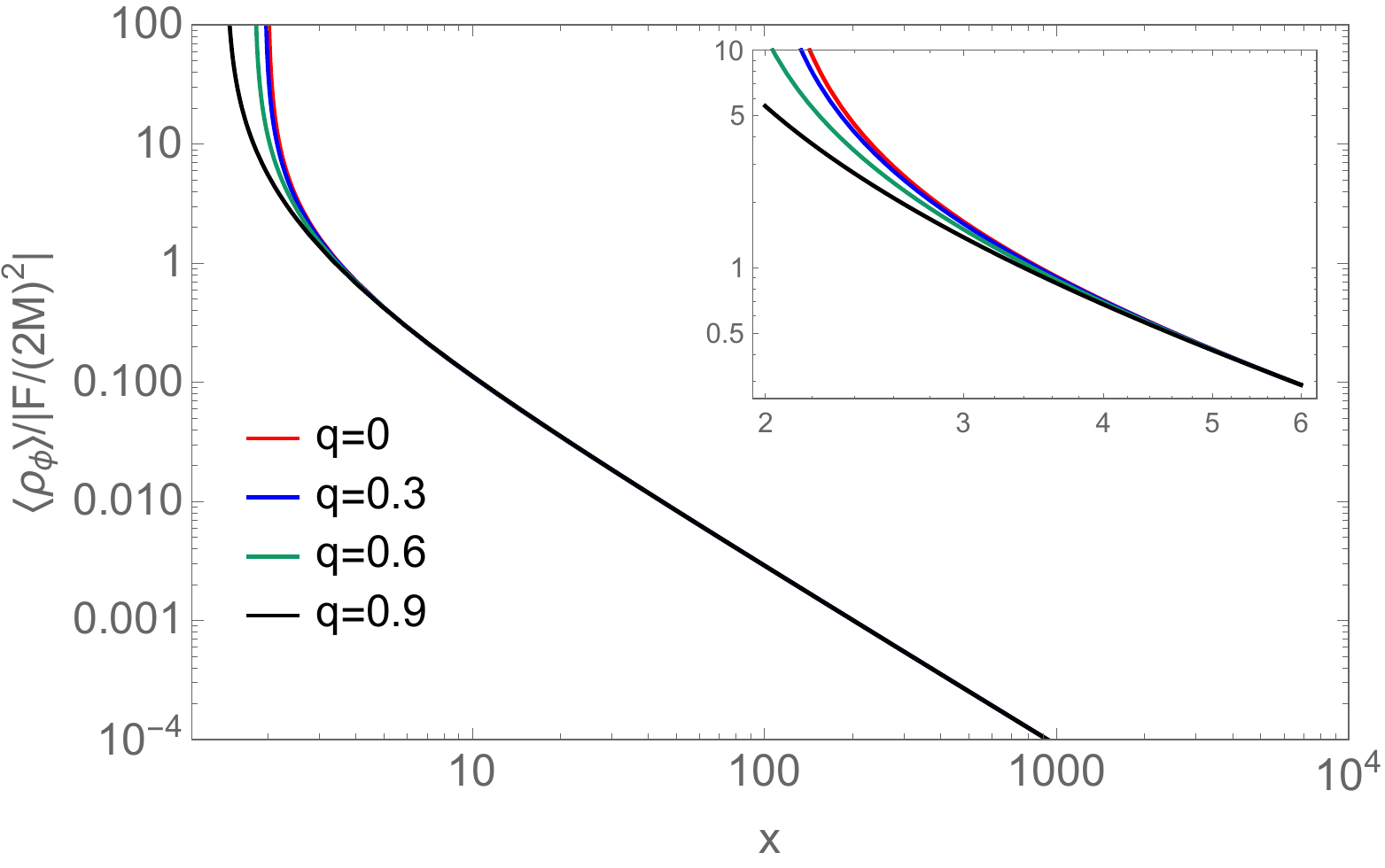}
    \caption{Normalized energy density of the scalar field $\langle \rho_{\phi} \rangle / |F/(2M)^2|$ as a function of the radial coordinate x, for different charge-mass ratios q, as given by Eq. (\ref{eq. rho promedio SF libre}). The abrupt density spike around the respective horizon is a result of employing nonregular coordinates at the horizon. The zoomed-in window pertains to the region between the radius horizon $x_{\text{h}}$ and the innermost stable circular orbit $x_{\text{isco}}$.}
    \label{Densidad de energia SF libre}
\end{figure}

To provide a clearer understanding of the behavior of the massive SF around an RN-BH, Fig. (\ref{Densidad de energia SF libre}) displays the normalized energy density of the SF, referred to as $\langle \rho_{\phi} \rangle / |F/(2M)^2|$, for various values of $q$. It can be observed that at larger distances, the influence of $q$ becomes imperceptible. This fact arises,  not surprisingly,  because the term $(q/x)^2$ within the metric function rapidly diminishes with increasing distance. However, the zoomed-in window suggests that $q$ has an impact on the region located between the photon sphere $x_{\text{ph}}$ and the innermost stable circular orbit $x_{\text{isco}}$—a region where marginally bound orbits $x_{\text{mb}}$ are typically located, making it inevitable to fall into the RN-BH. As the charge is turned on, the density profile becomes less cuspy compared to the Schwarzschild case at small radii. At such distances, the profile no longer follows the simple scaling: $\rho \propto x^{-3/2}$. This is mainly because as $q$ increases, the horizon radius becomes smaller, reducing its effective cross-section for capturing DM particles. The region mentioned above is particularly of interest for current and upcoming BHs experiments in the strong field regime that could potentially reveal deviations from standard geometries, shedding light on ``scalar hair'' phenomena.

\subsubsection{Accretion}
\label{accretion}
\begin{figure}
    \centering
    \includegraphics[scale=0.335]{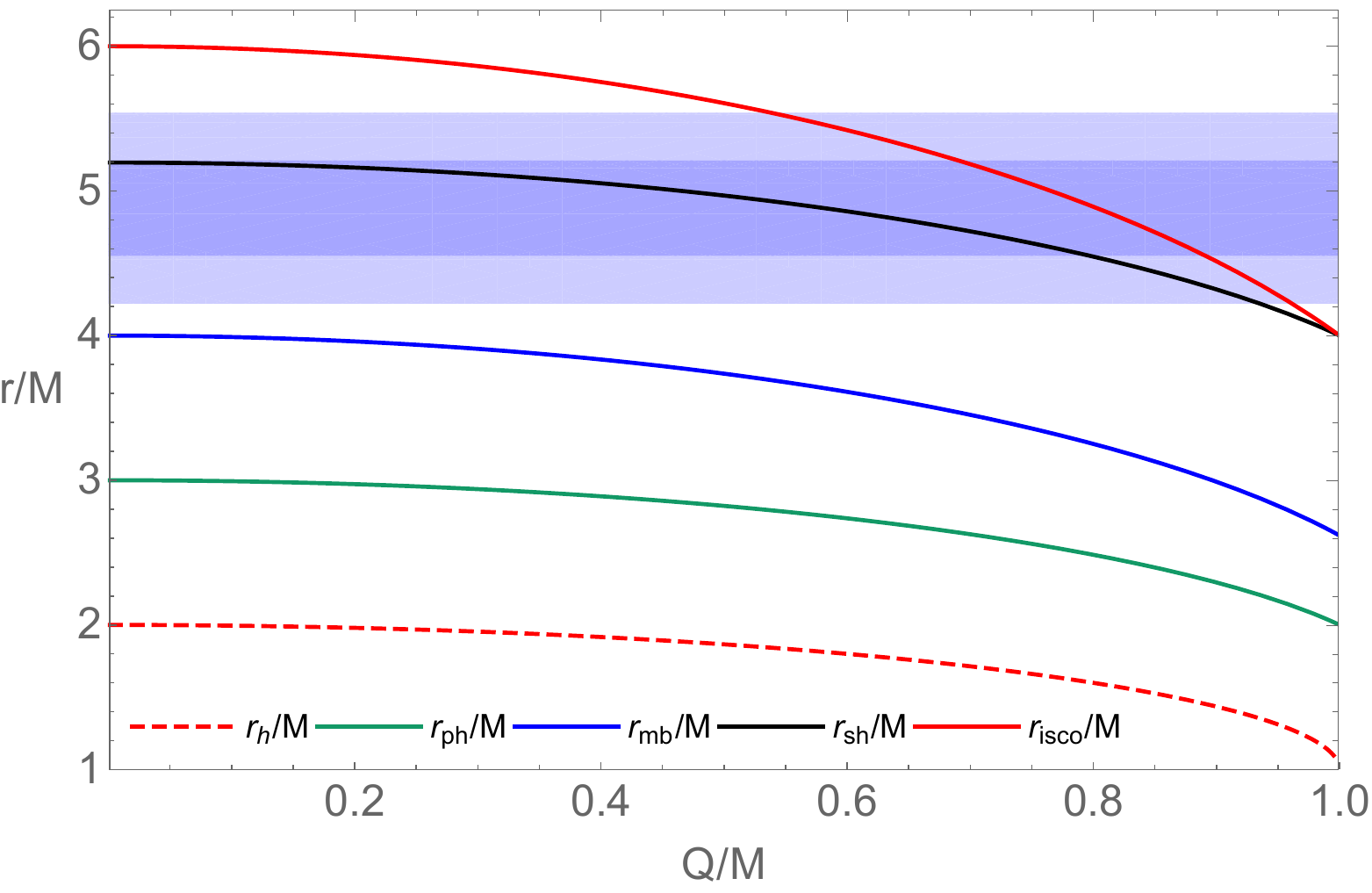}
    \caption{The image displays various orbits and regions of interest in the vicinity of a RN-BH. We can distinguish the horizon $r_{\text{h}}/M$ (dashed red line) \cite{Pugliese:2010ps}, the photon sphere $r_{\text{ph}}/M$ (Green line) \cite{Pugliese:2010ps,Claudel:2000yi}, the marginally bound radius $r_{\text{mb}}/M$ (blue line) \cite{Beheshti:2015bak}, the shadow radius $r_{\text{sh}}/M$ of the RN-BH (black line) \cite{Vagnozzi:2022moj}, and the innermost stable circular orbit $r_{\text{isco}}/M$ (red line) \cite{Pugliese:2010ps,Gomez:2023wei}. Additionally, the sky blue region $ 4.55 \lesssim r_{\text{sh}}/M \lesssim 5.21 $ corresponds to the $1\sigma$ constraint, while the light blue region $ 4.21 \lesssim r_{\text{sh}}/M \lesssim 5.56 $ corresponds to the $2\sigma$ constraint, set by Keck and VLTI telescopes, respectively \cite{Vagnozzi:2022moj}. The EHT provides the constraint $ Q/M \in [0,0.9] $ \cite{EventHorizonTelescope:2021dqv,EventHorizonTelescope:2021srq,EventHorizonTelescope:2022xqj,Vagnozzi:2022moj}.}
    \label{orbitas de interes}
\end{figure}

We compute the mass accretion rate around a RN-BH. The mass accretion can be obtained from the continuity equation is associated with the component $\nu = t$ of the
conservation equations $\nabla_\mu T^\mu_\nu = 0$. For a steady state, the mass accretion rate is defined as the energy flow through a closed surface of a sphere, given by
\begin{equation}
    \dot M(r) = \oint T^r_t\sqrt{-g}d\theta d\varphi,\label{definicion de M punto}
\end{equation}
where
\begin{equation}
    T_t^r = \frac{1}{g}\frac{\partial \phi}{\partial r}\frac{\partial \phi}{\partial t}.
\end{equation}
Using the SF profile (\ref{eq. SF real Maledung}) and considering the large-mass limit, we can calculate the mass accretion of DM\footnote{Notice that the mass accretion rate does not depend on the initial DM distribution because the mass that has fallen into the BH over time is balanced by the decrease of the DM distribution. In this sense, we can choose to work from the current time ($t=0$), which represents the remaining mass in the scalar cloud.}
\begin{equation}
    \dot M_{\text{SFDM}} = 4\pi r^2 \rho \sqrt{1-f},
    \label{eq. M punto SFDM}
\end{equation}
We can observe that the above expression is regular at the horizon. Typically, $r_{\text{isco}}$ is identified as the region where the maximum accretion rate is experienced \cite{Shapiro:1983du,b61f022b-8347-33e4-bf2e-c3d95dd12274,2010A&A...521A..15A}. However, other studies \cite{Beheshti:2015bak,Ravanal:2023ytp,Hod:2013mgr} suggest that accretion occurs in the region where marginally bound orbits ($ r_{\text{mb}} \lesssim r \lesssim r_{\text{isco}}$) are located. Indeed, it is in this region where particle capture occurs. Hence, we shall restrict our analysis to such scales. To better visualize this situation, we show in Fig. (\ref{orbitas de interes}) various orbits of interest. In particular, we assume that the maximum rate is attained at $r_{\text{mb}}$. Our previous study \cite{Ravanal:2023ytp}, in the case of a self-interacting scalar field, supports this possibility by indicating that the maximum rate occurs on marginally bound orbits.

We observe a decrease of approximately 50\% and 42\% in the accretion rate for the maximum allowed charge $q=0.9$ at $r_{\text{mb}}$ and $r_{\text{isco}}$, respectively, compared to the uncharged case. To get an idea of the order of magnitude of accretion in this scenario, we considered a Milky Way-like galaxy. Some studies \cite{Sadeghian:2013laa,Gondolo:1999ef} suggest that the dark matter density $\rho_{\text{DM}}$ in the vicinity of these BHs could be of the order of $\rho_{\text{DM}} \thicksim 10^6 M_{\odot}\;\text{pc}^{-3}$. Based on this argument, we obtained an accretion rate of the order of $ \dot M_{\text{SFDM}} \thicksim 10^{-8} M_{\odot} \;\text{yr}^{-1}$ in our region of interest. It is important to note that this result is sensitive to $\rho_{\text{DM}}$, that is, the type of model being considered. However, this result is consistent with our previous research \cite{Ravanal:2023ytp}, in which we found that in the repulsive SIDM model, the accretion obtained was of the order of $\dot M_{\rm SIDM}\sim10^{-10}M_\odot~{\rm yr}^{-1}$ \cite{Feng:2021qkj,Ravanal:2023ytp}. This result makes sense, as the existence of repulsion between particles suggests that the number of particles falling into the BH should be lower. However, all these estimations are small compared to baryonic matter accretion, suggesting the possibility that gravitationally bound structures composed of DM remain present in the galaxy. In fact, these structures should have a critical mass that surpasses millions of solar masses to reach the Bondi-like pressure-regulated infall \cite{Gomez:2024ack}.

\section{Comparison with previous results}
\label{sec:III}

\subsubsection{Self-interacting scalar field dark matter}
\label{self interactive scalar}
We present a brief summary of our previous work \cite{Ravanal:2023ytp} and highlight how this new work connects with it. When a quartic self-interaction potential is included in Eq. (\ref{Equation motion}), the nonlinear Klein-Gordon equation is obtained. In the large scalar-mass limit (\ref{eq. limite masa grande}), this equation can be identified as a Duffing-type equation \cite{kovacic2011duffing}, which describes a nonlinear harmonic oscillator, and its analytical solution can be expressed in terms of Jacobi elliptic functions \cite{Brax:2019npi,Frasca:2009bc}. It is important to note that at distances beyond the transition radius $r_{\text{sg}}$, we enter the domain where the self-gravity of scalar cloud turns out to be important. This arises from the hydrostatic equilibrium between the repulsive self-interaction and the self-gravity of the scalar cloud, unlike in the FDM case where the balance occurs between the self-gravity of the scalar cloud and the quantum pressure arising from the uncertainty principle.

\begin{figure}
    \centering
    \includegraphics[scale=0.34]{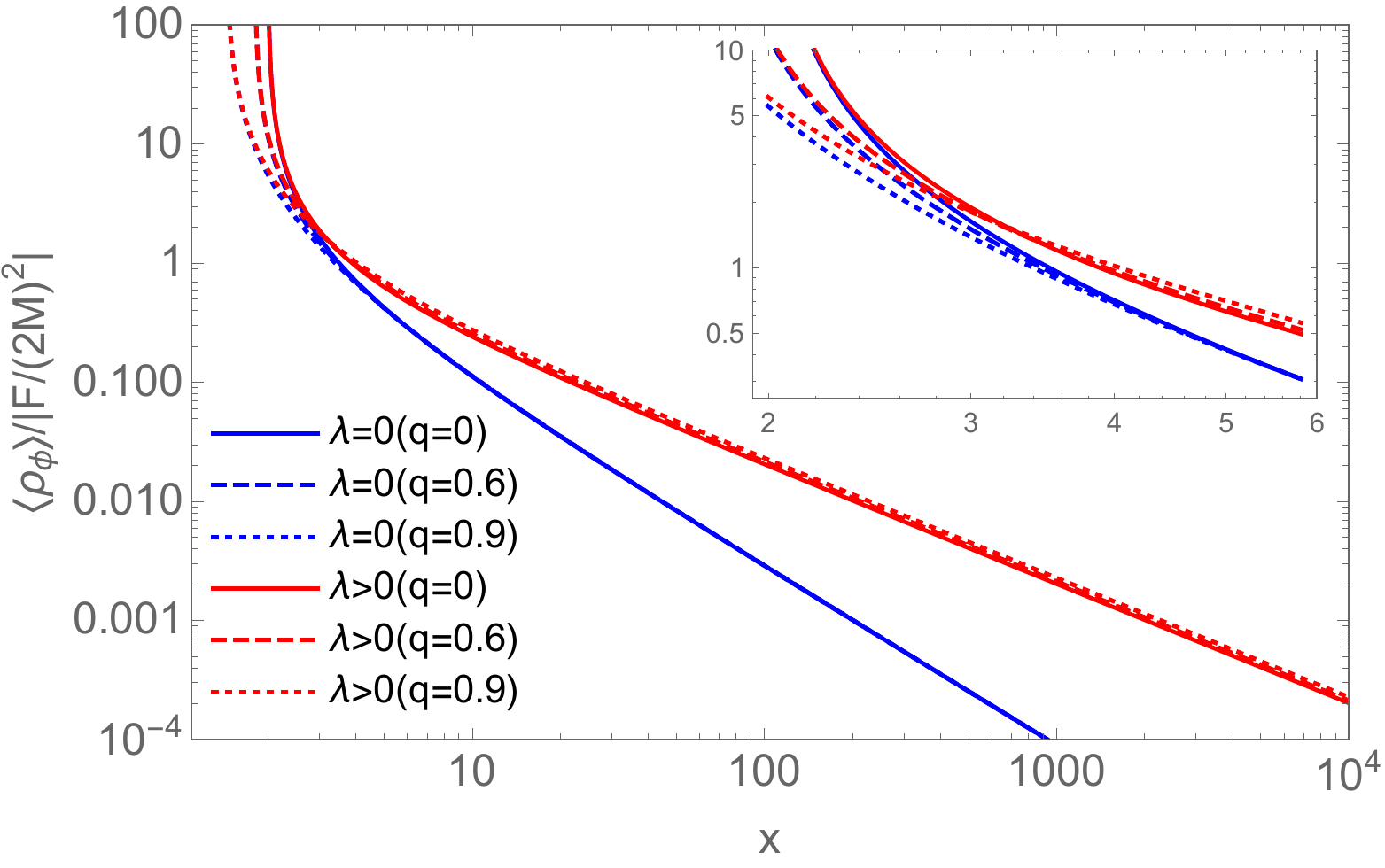}
    \caption{Comparison between the normalized energy density of the self-interacting scalar field  (see Eq (32) in \cite{Ravanal:2023ytp}) in red and the non-interacting one (\ref{eq. rho promedio SF libre}) in blue, as a function of the radial coordinate $x$ for $q = 0, 0.6, 0.9$. Here $\lambda$ represents the self-interaction of the SF. The small zoom corresponds to the region between $x_{\text{h}}$ and $x_{\text{isco}}$.}
    \label{comparativa}
\end{figure}

In Fig. (\ref{comparativa}), we present a comparison between the case of repulsive self-interaction (see Eq (32) in \cite{Ravanal:2023ytp}) and the non-interacting case (\ref{eq. rho promedio SF libre}). The first thing we notice is a significant change in slope between both cases. This difference lies in that the repulsive self-interaction stabilizes the self-gravity of the SF cloud. In the non-interacting scenario, the density decreases as $\langle \rho_{\phi} \rangle \propto r^{-3/2}$, while in the self-interacting scenario, it decreases as $\langle \rho_{\phi} \rangle \propto r^{-1}$, at radii between $r_{\rm h} \ll r \ll r_{\rm sg}$.\footnote{The density profile in the self-interacting case exhibits a behavior similar to the NFW profile $\rho_{\text{NFW}} \propto r^{-1}$ \cite{Navarro:1995iw} on scales smaller than the transition radius. This behavior has also been observed in previous studies \cite{Brax:2019npi,DeLuca:2023laa}. We remind that the change in slope for $\rho_{\text{NFW}}$ occurs around 1 kpc or less in Milky Way-like galaxies.} Such behaviors strictly hold when the effect of the charge becomes unimportant, which happens at radii larger than the $x_{isco}$, as can be seen in the figure. 

More importantly, in the small zoom of the image, we can see the impact of $q$ in the regions delimited by $x_{\text{ph}}$ and $x_{\text{isco}}$. This region is known as the marginally bound orbit $x_{\text{mb}}$, where the fall into the BH is inevitable for any massive particle. As we explained earlier, this fact is linked to the decrease in the cross-section due to the modification of the horizon caused by the effect of $q$. At small radii, and as the charge increases in both cases, the density profile becomes then less cuspy compared to the Schwarzschild case, as can be inferred from Eq. (\ref{eq. rho promedio SF libre}). As a consequence of the charge, the profile no longer follows the simple scaling found in the uncharged case: $\rho \propto r^{-3/2}$ or $\rho \propto r^{-1}$ at small radii.

On the other hand, we notice that both curves approach each other as we get closer to the BH. The reason for this is that self-interactions cannot counteract the gravitational effects near the horizon (see Appendix in \cite{Ravanal:2023ytp} for an explicit demonstration). This is because in the strong field regime, self-interactions become negligible. In the opposite scale, we observe that the effect of the charge is practically negligible, as the metric function decreases as $(q/x)^2$ at large distances. The effect of the charge is barely appreciable for the self-interacting case.


As a final remark, we have verified that the radial velocities also differ, being $v_r \propto r^{-1}$ \cite{Brax:2019npi} and $v_r \propto r^{-1/2}$ (see Eq. (\ref{eq.vr schw})) for the interactive and non-interactive cases, respectively. This is due to the need to satisfy the condition of constant energy flux, that is, independent of radii $r$.

\subsubsection{Axion-like dark matter particles}
\label{ULDM}

As a proof of concept, we compare the SF profile in different particle mass regimes, covering a wide class of SF dark matter scenarios. We pay particular attention to the large mass limit. Unfortunately, since the charged case has not been examined in such cases, we must restrict this comparison to the Schwarzschild BH. Within our approach, such a comparison makes sense at large radii when the term $Q^2/r^2$ can be safely ignored. Several authors have previously examined the massive Klein-Gordon equation \cite{Unruh:1976fm,Detweiler:1980uk}. The exact solutions to this equation can be expressed in terms of the confluent Heun function \cite{Fiziev:2005ki,Fiziev:2006tx,Bezerra:2013iha,Vieira:2014waa,Konoplya:2006br,Barranco:2012qs}. This problem has been revisited by \cite{Hui:2019aqm} and recently extended by \cite{Bucciotti:2023bvw}, including the effects of angular momentum in the SF.

In the context of a BH immersed in DM, there is a possibility of developing a ``scalar hair''. For a  massive and oscillating non-interacting SF, the authors of \cite{Hui:2019aqm} have identified different mass regimes (see Table 1 in \cite{Hui:2019aqm}) that can distinctly impact the scalar field profile. Regime IV corresponds to $ m > r_{\text{h}}^{-1}$, while Regime III is identified with $(r_{\text{sg}}r_{\text{h}})^{-1/2} \lesssim m \lesssim r_{\text{h}}^{-1}$. Furthermore, they divided the surroundings of the BH into two clearly defined regions: the first region is dominated by the BH's geometry, extending up to the transition radius $r_{\text{sg}}$, while the second region lies beyond this point. Our large-mass regime ($ r_{\text{h}} \gg 1/m $) corresponds indeed to the particle limit (regime IV in their prescription). Accordingly, they obtained the following expression for the SF profile
\begin{equation}
    r_{\text{h}} \lesssim r \lesssim r_{\text{sg}} \; \; : \; \; \phi \thicksim r^{-3/4} \mathrm{e}^{-imt} \mathrm{e}^{-i2m\sqrt{r r_h}}.
\end{equation}
We can observe that the amplitude of the SF profile behaves in the same way as in our case Eq. (\ref{eq.profile SF}), as it follows the same power law $\phi \propto r^{-3/4}$. The regime III is particularly interesting because it encompasses both particle-like and wave-like behavior
\begin{equation}
m^{-2}r_{\text{h}}^{-1} \lesssim r \lesssim r_{\text{sg}} : \phi \thicksim r^{-3/4} \mathrm{e}^{-imt} \cos \: (2m \sqrt{r r_h} - 3 \pi/4).    
\end{equation}
In fact, its amplitude exhibits the same behavior as $\phi \propto r^{-3/4}$, but with an additional modulation. They obtained a density profile $\rho \propto r^{-3/2}$. This profile is interpreted as a constant and steady energy flow falling into the BH at a velocity $v_r \propto r^{-1/2}$, which aligns with the behavior identified in Eq. (\ref{eq. rho promedio SF libre}). As an aside, in the regime of small mass, the aforementioned author asserts that the SF profile behaves according to the power law $\phi \propto r^{-1}$, as in the Jacobson's preliminary result \cite{Jacobson:1999vr}.

As noted in \cite{Bucciotti:2023bvw}, in the large mass regime and for distances far from the BH, the SF profile is independent of the angular momentum, as it follows the same power law $\phi \propto r^{-3/4}$ (see Eq. (2.19) in \cite{Bucciotti:2023bvw}). In the same direction, the authors of \cite{Clough:2019jpm} numerically calculated the SF profile in the large mass regime, considering the backreaction of the SF, and found that at significant distances from the BH, the envelopes of the SF profile follow the same power law $\phi \propto r^{-3/4}$ (see Fig. 1 in \cite{Clough:2019jpm}).

\section{DISCUSSION AND CONCLUSION}
\label{sec:IV}

The last decade has been thrilling in terms of astrophysical observations, reaching unprecedented resolutions, even on the order of the event horizon. These observations, conducted in the strong field regime, have the potential to provide unique insights not only into the intrinsic properties of BHs but also into the environments they inhabit. A key assumption in this context is the potential influence of DM on these observations, leaving distinctive signatures that could contribute to unveiling its elusive nature.

Therefore, it is of vital importance to develop models that deviate from the Schwarzschild paradigm. In this line of thinking, our study is situated, addressing the non-interactive case of the SF model and exploring the impact of electric charge $Q$ on the behavior of SF around a RN-BH. All of this is coupled with the results obtained by the EHT collaboration, which leaves open the possibility that BHs may have a non-zero electric charge. In general terms, with a 2$\sigma$ confidence level, this electric charge is found in an approximate range of $ q = Q/M \in [0,0.9]$. Therefore, we limit ourselves to considering these constraints and connect these results with our previous research \cite{Ravanal:2023ytp}.

The effect of $Q$ becomes relevant in the vicinity of the BH, especially in the region where marginally bound orbits are located, as shown in Fig. (\ref{SF profile}) and Fig. (\ref{Densidad de energia SF libre}). As expected, this effect diminishes due to the $r^{-2}$ term present in the metric function $f$ and becomes less relevant as we move away from the RN-BH. At such distances, the scalar field profile decreases as $\phi\propto r^{-3/4}$, while the density profile decreases as $\rho\propto r^{-3/2}$ and the radial velocity as $v_{r}\propto r^{-1/2}$.

A notable fact is the change in slope in the density profile between the non-interactive and interactive cases, with the behavior of the latter being proportional to $r^{-1}$, and its radial velocity also following suit. This change is due to the nature of the fields, as the repulsive self-interaction slows down the fall of DM. Additionally, we require a steady-state behavior for the flow of the fields, so it must be independent of $r$. These values fall within the range typically considered by different models of power-law exponents for DM density profiles $\rho_{\text{DM}} \propto r^{- \gamma}$ for $\gamma$ $\in [0,5]$ \cite{Ferreira:2020fam}.

An interesting inference is uncovered from Fig. (\ref{comparativa}): the charge has a non-trivial impact on the density profile within marginally bound orbits. This leads us to the question: is it possible to extract information about the DM properties from BH observations within a beyond the Schwarzschild geometry? At first glance, it seems very challenging, but with the upcoming high-resolution BH experiments in the strong-field scale, we expect to detect the famous ``scalar hair'' and characterize the spacetime geometry unprecedentedly.


We found that the maximum accretion rate of the SF in marginally bound orbits, i.e., between $r_{\text{mb}}$ and $r_{\text{isco}}$, decreases by up to 50 \% in the case of maximum allowed charge compared to the uncharged scenario. The choice of $r_{\text{mb}}$ as the upper limit is based on the premise that, with the SF having a nonzero mass, the capture of particles in these orbits is more probable and the fall into the BH seems inevitable, as there is no point of return once entering that radius. We also obtained an order of magnitude estimate for the accretion of SF dark matter $\dot M_{\text{SFDM}} \thicksim 10^{-8} M_{\odot} \;\text{yr}^{-1}$, which is higher compared to the self-interacting case $\dot M_{\rm SIDM}\sim10^{-10}M_\odot~{\rm yr}^{-1}$ \cite{Ravanal:2023ytp}. In both cases, the value is small compared to the usual accretion of baryons (Bondi or Eddington) reported in previous research \cite{Salpeter:1964kb,Bondi:1944jm,Bondi:1952ni,Edgar:2004mk,EventHorizonTelescope:2021srq}. This relatively small number leaves open the possibility that structures at the subgalactic or galactic level composed of DM continue to exist at present times, as these structures are typically estimated to have hundreds of thousands or millions of solar masses.

In summary, current and future observations of BHs provide a unique opportunity to investigate the role that DM plays in the dynamics of these cosmic objects and their surroundings. These observations, supported by theoretical advances, have the potential to shed light on the nature of DM and its interactions with BHs. Although our study is modest, it has provided a different perspective, often overlooked, on the impact of electric charge on BHs and its secular effect in the vicinity. We hope that our findings open new avenues for future research in this fascinating field, contributing to the understanding of this elusive component of our Universe.

\section{ACKNOWLEDGEMENTS}
Y. R is supported by Beca Doctorado Convenio Marco de la Universidad de Santiago de Chile (USACH) from 2019-2020 and by Beca Doctorado Nacional año 2021 Folio  No. 21211644 de la Agencia Nacional de Investigacion y Desarrollo de Chile (ANID). Y. R also acknowledges Fernando M\'endez for his guidance in this process. G. G acknowledges financial support from Agencia Nacional de Investigaci\'on y Desarrollo (ANID), Chile, through the FONDECYT postdoctoral Grant No. 3210417. N.C acknowledges the support of Universidad de Santiago de Chile (USACH), through Proyecto DICYT N° 042131CM, Vicerrectoría de Investigación, Desarrollo e Innovación. We thank the anonymous referees for their valuable comments and suggestions.

\appendix*
\label{Apendice}

\bibliography{sample}

\begin{thebibliography}{131}%
\makeatletter
\providecommand \@ifxundefined [1]{%
 \@ifx{#1\undefined}
}%
\providecommand \@ifnum [1]{%
 \ifnum #1\expandafter \@firstoftwo
 \else \expandafter \@secondoftwo
 \fi
}%
\providecommand \@ifx [1]{%
 \ifx #1\expandafter \@firstoftwo
 \else \expandafter \@secondoftwo
 \fi
}%
\providecommand \natexlab [1]{#1}%
\providecommand \enquote  [1]{``#1''}%
\providecommand \bibnamefont  [1]{#1}%
\providecommand \bibfnamefont [1]{#1}%
\providecommand \citenamefont [1]{#1}%
\providecommand \href@noop [0]{\@secondoftwo}%
\providecommand \href [0]{\begingroup \@sanitize@url \@href}%
\providecommand \@href[1]{\@@startlink{#1}\@@href}%
\providecommand \@@href[1]{\endgroup#1\@@endlink}%
\providecommand \@sanitize@url [0]{\catcode `\\12\catcode `\$12\catcode
  `\&12\catcode `\#12\catcode `\^12\catcode `\_12\catcode `\%12\relax}%
\providecommand \@@startlink[1]{}%
\providecommand \@@endlink[0]{}%
\providecommand \url  [0]{\begingroup\@sanitize@url \@url }%
\providecommand \@url [1]{\endgroup\@href {#1}{\urlprefix }}%
\providecommand \urlprefix  [0]{URL }%
\providecommand \Eprint [0]{\href }%
\providecommand \doibase [0]{http://dx.doi.org/}%
\providecommand \selectlanguage [0]{\@gobble}%
\providecommand \bibinfo  [0]{\@secondoftwo}%
\providecommand \bibfield  [0]{\@secondoftwo}%
\providecommand \translation [1]{[#1]}%
\providecommand \BibitemOpen [0]{}%
\providecommand \bibitemStop [0]{}%
\providecommand \bibitemNoStop [0]{.\EOS\space}%
\providecommand \EOS [0]{\spacefactor3000\relax}%
\providecommand \BibitemShut  [1]{\csname bibitem#1\endcsname}%
\let\auto@bib@innerbib\@empty
\bibitem [{\citenamefont {Spergel}\ \emph {et~al.}(2003)\citenamefont {Spergel}
  \emph {et~al.}}]{WMAP:2003elm}%
  \BibitemOpen
  \bibfield  {author} {\bibinfo {author} {\bibfnamefont {D.~N.}\ \bibnamefont
  {Spergel}} \emph {et~al.} (\bibinfo {collaboration} {WMAP}),\ }\href
  {\doibase 10.1086/377226} {\bibfield  {journal} {\bibinfo  {journal}
  {Astrophys. J. Suppl.}\ }\textbf {\bibinfo {volume} {148}},\ \bibinfo {pages}
  {175} (\bibinfo {year} {2003})},\ \Eprint
  {http://arxiv.org/abs/astro-ph/0302209} {arXiv:astro-ph/0302209} \BibitemShut
  {NoStop}%
\bibitem [{\citenamefont {Ade}\ \emph {et~al.}(2016)\citenamefont {Ade} \emph
  {et~al.}}]{Planck:2015fie}%
  \BibitemOpen
  \bibfield  {author} {\bibinfo {author} {\bibfnamefont {P.~A.~R.}\
  \bibnamefont {Ade}} \emph {et~al.} (\bibinfo {collaboration} {Planck}),\
  }\href {\doibase 10.1051/0004-6361/201525830} {\bibfield  {journal} {\bibinfo
   {journal} {Astron. Astrophys.}\ }\textbf {\bibinfo {volume} {594}},\
  \bibinfo {pages} {A13} (\bibinfo {year} {2016})},\ \Eprint
  {http://arxiv.org/abs/1502.01589} {arXiv:1502.01589 [astro-ph.CO]}
  \BibitemShut {NoStop}%
\bibitem [{\citenamefont {Aghanim}\ \emph {et~al.}(2020)\citenamefont {Aghanim}
  \emph {et~al.}}]{Planck:2018vyg}%
  \BibitemOpen
  \bibfield  {author} {\bibinfo {author} {\bibfnamefont {N.}~\bibnamefont
  {Aghanim}} \emph {et~al.} (\bibinfo {collaboration} {Planck}),\ }\href
  {\doibase 10.1051/0004-6361/201833910} {\bibfield  {journal} {\bibinfo
  {journal} {Astron. Astrophys.}\ }\textbf {\bibinfo {volume} {641}},\ \bibinfo
  {pages} {A6} (\bibinfo {year} {2020})},\ \bibinfo {note} {[Erratum:
  Astron.Astrophys. 652, C4 (2021)]},\ \Eprint
  {http://arxiv.org/abs/1807.06209} {arXiv:1807.06209 [astro-ph.CO]}
  \BibitemShut {NoStop}%
\bibitem [{\citenamefont {Tegmark}\ \emph {et~al.}(2004)\citenamefont {Tegmark}
  \emph {et~al.}}]{SDSS:2003tbn}%
  \BibitemOpen
  \bibfield  {author} {\bibinfo {author} {\bibfnamefont {M.}~\bibnamefont
  {Tegmark}} \emph {et~al.} (\bibinfo {collaboration} {SDSS}),\ }\href
  {\doibase 10.1086/382125} {\bibfield  {journal} {\bibinfo  {journal}
  {Astrophys. J.}\ }\textbf {\bibinfo {volume} {606}},\ \bibinfo {pages} {702}
  (\bibinfo {year} {2004})},\ \Eprint {http://arxiv.org/abs/astro-ph/0310725}
  {arXiv:astro-ph/0310725} \BibitemShut {NoStop}%
\bibitem [{\citenamefont {Anderson}\ \emph {et~al.}(2014)\citenamefont
  {Anderson} \emph {et~al.}}]{BOSS:2013rlg}%
  \BibitemOpen
  \bibfield  {author} {\bibinfo {author} {\bibfnamefont {L.}~\bibnamefont
  {Anderson}} \emph {et~al.} (\bibinfo {collaboration} {BOSS}),\ }\href
  {\doibase 10.1093/mnras/stu523} {\bibfield  {journal} {\bibinfo  {journal}
  {Mon. Not. Roy. Astron. Soc.}\ }\textbf {\bibinfo {volume} {441}},\ \bibinfo
  {pages} {24} (\bibinfo {year} {2014})},\ \Eprint
  {http://arxiv.org/abs/1312.4877} {arXiv:1312.4877 [astro-ph.CO]} \BibitemShut
  {NoStop}%
\bibitem [{\citenamefont {Alam}\ \emph {et~al.}(2017)\citenamefont {Alam} \emph
  {et~al.}}]{BOSS:2016wmc}%
  \BibitemOpen
  \bibfield  {author} {\bibinfo {author} {\bibfnamefont {S.}~\bibnamefont
  {Alam}} \emph {et~al.} (\bibinfo {collaboration} {BOSS}),\ }\href {\doibase
  10.1093/mnras/stx721} {\bibfield  {journal} {\bibinfo  {journal} {Mon. Not.
  Roy. Astron. Soc.}\ }\textbf {\bibinfo {volume} {470}},\ \bibinfo {pages}
  {2617} (\bibinfo {year} {2017})},\ \Eprint {http://arxiv.org/abs/1607.03155}
  {arXiv:1607.03155 [astro-ph.CO]} \BibitemShut {NoStop}%
\bibitem [{\citenamefont {Navarro}\ \emph {et~al.}(1996)\citenamefont
  {Navarro}, \citenamefont {Frenk},\ and\ \citenamefont
  {White}}]{Navarro:1995iw}%
  \BibitemOpen
  \bibfield  {author} {\bibinfo {author} {\bibfnamefont {J.~F.}\ \bibnamefont
  {Navarro}}, \bibinfo {author} {\bibfnamefont {C.~S.}\ \bibnamefont {Frenk}},
  \ and\ \bibinfo {author} {\bibfnamefont {S.~D.~M.}\ \bibnamefont {White}},\
  }\href {\doibase 10.1086/177173} {\bibfield  {journal} {\bibinfo  {journal}
  {Astrophys. J.}\ }\textbf {\bibinfo {volume} {462}},\ \bibinfo {pages} {563}
  (\bibinfo {year} {1996})},\ \Eprint {http://arxiv.org/abs/astro-ph/9508025}
  {arXiv:astro-ph/9508025} \BibitemShut {NoStop}%
\bibitem [{\citenamefont {Moore}(1994)}]{Moore:1994yx}%
  \BibitemOpen
  \bibfield  {author} {\bibinfo {author} {\bibfnamefont {B.}~\bibnamefont
  {Moore}},\ }\href {\doibase 10.1038/370629a0} {\bibfield  {journal} {\bibinfo
   {journal} {Nature}\ }\textbf {\bibinfo {volume} {370}},\ \bibinfo {pages}
  {629} (\bibinfo {year} {1994})}\BibitemShut {NoStop}%
\bibitem [{\citenamefont {de~Blok}(2010)}]{deBlok:2009sp}%
  \BibitemOpen
  \bibfield  {author} {\bibinfo {author} {\bibfnamefont {W.~J.~G.}\
  \bibnamefont {de~Blok}},\ }\href {\doibase 10.1155/2010/789293} {\bibfield
  {journal} {\bibinfo  {journal} {Adv. Astron.}\ }\textbf {\bibinfo {volume}
  {2010}},\ \bibinfo {pages} {789293} (\bibinfo {year} {2010})},\ \Eprint
  {http://arxiv.org/abs/0910.3538} {arXiv:0910.3538 [astro-ph.CO]} \BibitemShut
  {NoStop}%
\bibitem [{\citenamefont {Oh}\ \emph {et~al.}(2011)\citenamefont {Oh},
  \citenamefont {Brook}, \citenamefont {Governato}, \citenamefont {Brinks},
  \citenamefont {Mayer}, \citenamefont {de~Blok}, \citenamefont {Brooks},\ and\
  \citenamefont {Walter}}]{Oh:2010mc}%
  \BibitemOpen
  \bibfield  {author} {\bibinfo {author} {\bibfnamefont {S.-H.}\ \bibnamefont
  {Oh}}, \bibinfo {author} {\bibfnamefont {C.}~\bibnamefont {Brook}}, \bibinfo
  {author} {\bibfnamefont {F.}~\bibnamefont {Governato}}, \bibinfo {author}
  {\bibfnamefont {E.}~\bibnamefont {Brinks}}, \bibinfo {author} {\bibfnamefont
  {L.}~\bibnamefont {Mayer}}, \bibinfo {author} {\bibfnamefont {W.~J.~G.}\
  \bibnamefont {de~Blok}}, \bibinfo {author} {\bibfnamefont {A.}~\bibnamefont
  {Brooks}}, \ and\ \bibinfo {author} {\bibfnamefont {F.}~\bibnamefont
  {Walter}},\ }\href {\doibase 10.1088/0004-6256/142/1/24} {\bibfield
  {journal} {\bibinfo  {journal} {Astron. J.}\ }\textbf {\bibinfo {volume}
  {142}},\ \bibinfo {pages} {24} (\bibinfo {year} {2011})},\ \Eprint
  {http://arxiv.org/abs/1011.2777} {arXiv:1011.2777 [astro-ph.CO]} \BibitemShut
  {NoStop}%
\bibitem [{\citenamefont {Moore}\ \emph {et~al.}(1999)\citenamefont {Moore},
  \citenamefont {Ghigna}, \citenamefont {Governato}, \citenamefont {Lake},
  \citenamefont {Quinn}, \citenamefont {Stadel},\ and\ \citenamefont
  {Tozzi}}]{Moore:1999nt}%
  \BibitemOpen
  \bibfield  {author} {\bibinfo {author} {\bibfnamefont {B.}~\bibnamefont
  {Moore}}, \bibinfo {author} {\bibfnamefont {S.}~\bibnamefont {Ghigna}},
  \bibinfo {author} {\bibfnamefont {F.}~\bibnamefont {Governato}}, \bibinfo
  {author} {\bibfnamefont {G.}~\bibnamefont {Lake}}, \bibinfo {author}
  {\bibfnamefont {T.~R.}\ \bibnamefont {Quinn}}, \bibinfo {author}
  {\bibfnamefont {J.}~\bibnamefont {Stadel}}, \ and\ \bibinfo {author}
  {\bibfnamefont {P.}~\bibnamefont {Tozzi}},\ }\href {\doibase 10.1086/312287}
  {\bibfield  {journal} {\bibinfo  {journal} {Astrophys. J. Lett.}\ }\textbf
  {\bibinfo {volume} {524}},\ \bibinfo {pages} {L19} (\bibinfo {year}
  {1999})},\ \Eprint {http://arxiv.org/abs/astro-ph/9907411}
  {arXiv:astro-ph/9907411} \BibitemShut {NoStop}%
\bibitem [{\citenamefont {Bullock}\ \emph {et~al.}(2000)\citenamefont
  {Bullock}, \citenamefont {Kravtsov},\ and\ \citenamefont
  {Weinberg}}]{Bullock:2000wn}%
  \BibitemOpen
  \bibfield  {author} {\bibinfo {author} {\bibfnamefont {J.~S.}\ \bibnamefont
  {Bullock}}, \bibinfo {author} {\bibfnamefont {A.~V.}\ \bibnamefont
  {Kravtsov}}, \ and\ \bibinfo {author} {\bibfnamefont {D.~H.}\ \bibnamefont
  {Weinberg}},\ }\href {\doibase 10.1086/309279} {\bibfield  {journal}
  {\bibinfo  {journal} {Astrophys. J.}\ }\textbf {\bibinfo {volume} {539}},\
  \bibinfo {pages} {517} (\bibinfo {year} {2000})},\ \Eprint
  {http://arxiv.org/abs/astro-ph/0002214} {arXiv:astro-ph/0002214} \BibitemShut
  {NoStop}%
\bibitem [{\citenamefont {Boylan-Kolchin}\ \emph {et~al.}(2011)\citenamefont
  {Boylan-Kolchin}, \citenamefont {Bullock},\ and\ \citenamefont
  {Kaplinghat}}]{Boylan-Kolchin:2011qkt}%
  \BibitemOpen
  \bibfield  {author} {\bibinfo {author} {\bibfnamefont {M.}~\bibnamefont
  {Boylan-Kolchin}}, \bibinfo {author} {\bibfnamefont {J.~S.}\ \bibnamefont
  {Bullock}}, \ and\ \bibinfo {author} {\bibfnamefont {M.}~\bibnamefont
  {Kaplinghat}},\ }\href {\doibase 10.1111/j.1745-3933.2011.01074.x} {\bibfield
   {journal} {\bibinfo  {journal} {Mon. Not. Roy. Astron. Soc.}\ }\textbf
  {\bibinfo {volume} {415}},\ \bibinfo {pages} {L40} (\bibinfo {year}
  {2011})},\ \Eprint {http://arxiv.org/abs/1103.0007} {arXiv:1103.0007
  [astro-ph.CO]} \BibitemShut {NoStop}%
\bibitem [{\citenamefont {Garrison-Kimmel}\ \emph {et~al.}(2014)\citenamefont
  {Garrison-Kimmel}, \citenamefont {Boylan-Kolchin}, \citenamefont {Bullock},\
  and\ \citenamefont {Kirby}}]{Garrison-Kimmel:2014vqa}%
  \BibitemOpen
  \bibfield  {author} {\bibinfo {author} {\bibfnamefont {S.}~\bibnamefont
  {Garrison-Kimmel}}, \bibinfo {author} {\bibfnamefont {M.}~\bibnamefont
  {Boylan-Kolchin}}, \bibinfo {author} {\bibfnamefont {J.~S.}\ \bibnamefont
  {Bullock}}, \ and\ \bibinfo {author} {\bibfnamefont {E.~N.}\ \bibnamefont
  {Kirby}},\ }\href {\doibase 10.1093/mnras/stu1477} {\bibfield  {journal}
  {\bibinfo  {journal} {Mon. Not. Roy. Astron. Soc.}\ }\textbf {\bibinfo
  {volume} {444}},\ \bibinfo {pages} {222} (\bibinfo {year} {2014})},\ \Eprint
  {http://arxiv.org/abs/1404.5313} {arXiv:1404.5313 [astro-ph.GA]} \BibitemShut
  {NoStop}%
\bibitem [{\citenamefont {Del~Popolo}\ and\ \citenamefont
  {Le~Delliou}(2017)}]{DelPopolo:2016emo}%
  \BibitemOpen
  \bibfield  {author} {\bibinfo {author} {\bibfnamefont {A.}~\bibnamefont
  {Del~Popolo}}\ and\ \bibinfo {author} {\bibfnamefont {M.}~\bibnamefont
  {Le~Delliou}},\ }\href {\doibase 10.3390/galaxies5010017} {\bibfield
  {journal} {\bibinfo  {journal} {Galaxies}\ }\textbf {\bibinfo {volume} {5}},\
  \bibinfo {pages} {17} (\bibinfo {year} {2017})},\ \Eprint
  {http://arxiv.org/abs/1606.07790} {arXiv:1606.07790 [astro-ph.CO]}
  \BibitemShut {NoStop}%
\bibitem [{\citenamefont {Dutton}\ \emph {et~al.}(2019)\citenamefont {Dutton},
  \citenamefont {Macci\`o}, \citenamefont {Buck}, \citenamefont {Dixon},
  \citenamefont {Blank},\ and\ \citenamefont {Obreja}}]{Dutton:2018nop}%
  \BibitemOpen
  \bibfield  {author} {\bibinfo {author} {\bibfnamefont {A.~A.}\ \bibnamefont
  {Dutton}}, \bibinfo {author} {\bibfnamefont {A.~V.}\ \bibnamefont
  {Macci\`o}}, \bibinfo {author} {\bibfnamefont {T.}~\bibnamefont {Buck}},
  \bibinfo {author} {\bibfnamefont {K.~L.}\ \bibnamefont {Dixon}}, \bibinfo
  {author} {\bibfnamefont {M.}~\bibnamefont {Blank}}, \ and\ \bibinfo {author}
  {\bibfnamefont {A.}~\bibnamefont {Obreja}},\ }\href {\doibase
  10.1093/mnras/stz889} {\bibfield  {journal} {\bibinfo  {journal} {Mon. Not.
  Roy. Astron. Soc.}\ }\textbf {\bibinfo {volume} {486}},\ \bibinfo {pages}
  {655} (\bibinfo {year} {2019})},\ \Eprint {http://arxiv.org/abs/1811.10625}
  {arXiv:1811.10625 [astro-ph.GA]} \BibitemShut {NoStop}%
\bibitem [{\citenamefont {Dutton}\ \emph {et~al.}(2020)\citenamefont {Dutton},
  \citenamefont {Buck}, \citenamefont {Macci\`o}, \citenamefont {Dixon},
  \citenamefont {Blank},\ and\ \citenamefont {Obreja}}]{Dutton:2020vne}%
  \BibitemOpen
  \bibfield  {author} {\bibinfo {author} {\bibfnamefont {A.~A.}\ \bibnamefont
  {Dutton}}, \bibinfo {author} {\bibfnamefont {T.}~\bibnamefont {Buck}},
  \bibinfo {author} {\bibfnamefont {A.~V.}\ \bibnamefont {Macci\`o}}, \bibinfo
  {author} {\bibfnamefont {K.~L.}\ \bibnamefont {Dixon}}, \bibinfo {author}
  {\bibfnamefont {M.}~\bibnamefont {Blank}}, \ and\ \bibinfo {author}
  {\bibfnamefont {A.}~\bibnamefont {Obreja}},\ }\href {\doibase
  10.1093/mnras/staa3028} {\  (\bibinfo {year} {2020}),\
  10.1093/mnras/staa3028},\ \Eprint {http://arxiv.org/abs/2011.11351}
  {arXiv:2011.11351 [astro-ph.GA]} \BibitemShut {NoStop}%
\bibitem [{\citenamefont {Sommer-Larsen}\ and\ \citenamefont
  {Vedel}(1999)}]{Sommer-Larsen:1998ppj}%
  \BibitemOpen
  \bibfield  {author} {\bibinfo {author} {\bibfnamefont {J.}~\bibnamefont
  {Sommer-Larsen}}\ and\ \bibinfo {author} {\bibfnamefont {H.}~\bibnamefont
  {Vedel}},\ }\href {\doibase 10.1086/307374} {\bibfield  {journal} {\bibinfo
  {journal} {Astrophys. J.}\ }\textbf {\bibinfo {volume} {519}},\ \bibinfo
  {pages} {501} (\bibinfo {year} {1999})},\ \Eprint
  {http://arxiv.org/abs/astro-ph/9801094} {arXiv:astro-ph/9801094} \BibitemShut
  {NoStop}%
\bibitem [{\citenamefont {Milgrom}(1983)}]{Milgrom:1983ca}%
  \BibitemOpen
  \bibfield  {author} {\bibinfo {author} {\bibfnamefont {M.}~\bibnamefont
  {Milgrom}},\ }\href {\doibase 10.1086/161130} {\bibfield  {journal} {\bibinfo
   {journal} {Astrophys. J.}\ }\textbf {\bibinfo {volume} {270}},\ \bibinfo
  {pages} {365} (\bibinfo {year} {1983})}\BibitemShut {NoStop}%
\bibitem [{\citenamefont {Hu}\ \emph {et~al.}(2000)\citenamefont {Hu},
  \citenamefont {Barkana},\ and\ \citenamefont {Gruzinov}}]{Hu:2000ke}%
  \BibitemOpen
  \bibfield  {author} {\bibinfo {author} {\bibfnamefont {W.}~\bibnamefont
  {Hu}}, \bibinfo {author} {\bibfnamefont {R.}~\bibnamefont {Barkana}}, \ and\
  \bibinfo {author} {\bibfnamefont {A.}~\bibnamefont {Gruzinov}},\ }\href
  {\doibase 10.1103/PhysRevLett.85.1158} {\bibfield  {journal} {\bibinfo
  {journal} {Phys. Rev. Lett.}\ }\textbf {\bibinfo {volume} {85}},\ \bibinfo
  {pages} {1158} (\bibinfo {year} {2000})},\ \Eprint
  {http://arxiv.org/abs/astro-ph/0003365} {arXiv:astro-ph/0003365} \BibitemShut
  {NoStop}%
\bibitem [{\citenamefont {Hui}\ \emph {et~al.}(2017)\citenamefont {Hui},
  \citenamefont {Ostriker}, \citenamefont {Tremaine},\ and\ \citenamefont
  {Witten}}]{Hui:2016ltb}%
  \BibitemOpen
  \bibfield  {author} {\bibinfo {author} {\bibfnamefont {L.}~\bibnamefont
  {Hui}}, \bibinfo {author} {\bibfnamefont {J.~P.}\ \bibnamefont {Ostriker}},
  \bibinfo {author} {\bibfnamefont {S.}~\bibnamefont {Tremaine}}, \ and\
  \bibinfo {author} {\bibfnamefont {E.}~\bibnamefont {Witten}},\ }\href
  {\doibase 10.1103/PhysRevD.95.043541} {\bibfield  {journal} {\bibinfo
  {journal} {Phys. Rev. D}\ }\textbf {\bibinfo {volume} {95}},\ \bibinfo
  {pages} {043541} (\bibinfo {year} {2017})},\ \Eprint
  {http://arxiv.org/abs/1610.08297} {arXiv:1610.08297 [astro-ph.CO]}
  \BibitemShut {NoStop}%
\bibitem [{\citenamefont {Ferreira}(2021)}]{Ferreira:2020fam}%
  \BibitemOpen
  \bibfield  {author} {\bibinfo {author} {\bibfnamefont {E.~G.~M.}\
  \bibnamefont {Ferreira}},\ }\href {\doibase 10.1007/s00159-021-00135-6}
  {\bibfield  {journal} {\bibinfo  {journal} {Astron. Astrophys. Rev.}\
  }\textbf {\bibinfo {volume} {29}},\ \bibinfo {pages} {7} (\bibinfo {year}
  {2021})},\ \Eprint {http://arxiv.org/abs/2005.03254} {arXiv:2005.03254
  [astro-ph.CO]} \BibitemShut {NoStop}%
\bibitem [{\citenamefont {Sin}(1994)}]{Sin:1992bg}%
  \BibitemOpen
  \bibfield  {author} {\bibinfo {author} {\bibfnamefont {S.-J.}\ \bibnamefont
  {Sin}},\ }\href {\doibase 10.1103/PhysRevD.50.3650} {\bibfield  {journal}
  {\bibinfo  {journal} {Phys. Rev. D}\ }\textbf {\bibinfo {volume} {50}},\
  \bibinfo {pages} {3650} (\bibinfo {year} {1994})},\ \Eprint
  {http://arxiv.org/abs/hep-ph/9205208} {arXiv:hep-ph/9205208} \BibitemShut
  {NoStop}%
\bibitem [{\citenamefont {Ji}\ and\ \citenamefont {Sin}(1994)}]{Ji:1994xh}%
  \BibitemOpen
  \bibfield  {author} {\bibinfo {author} {\bibfnamefont {S.~U.}\ \bibnamefont
  {Ji}}\ and\ \bibinfo {author} {\bibfnamefont {S.~J.}\ \bibnamefont {Sin}},\
  }\href {\doibase 10.1103/PhysRevD.50.3655} {\bibfield  {journal} {\bibinfo
  {journal} {Phys. Rev. D}\ }\textbf {\bibinfo {volume} {50}},\ \bibinfo
  {pages} {3655} (\bibinfo {year} {1994})},\ \Eprint
  {http://arxiv.org/abs/hep-ph/9409267} {arXiv:hep-ph/9409267} \BibitemShut
  {NoStop}%
\bibitem [{\citenamefont {Seidel}\ and\ \citenamefont
  {Suen}(1990)}]{Seidel:1990jh}%
  \BibitemOpen
  \bibfield  {author} {\bibinfo {author} {\bibfnamefont {E.}~\bibnamefont
  {Seidel}}\ and\ \bibinfo {author} {\bibfnamefont {W.-M.}\ \bibnamefont
  {Suen}},\ }\href {\doibase 10.1103/PhysRevD.42.384} {\bibfield  {journal}
  {\bibinfo  {journal} {Phys. Rev. D}\ }\textbf {\bibinfo {volume} {42}},\
  \bibinfo {pages} {384} (\bibinfo {year} {1990})}\BibitemShut {NoStop}%
\bibitem [{\citenamefont {Matos}\ and\ \citenamefont
  {Guzman}(2000)}]{Matos:1998vk}%
  \BibitemOpen
  \bibfield  {author} {\bibinfo {author} {\bibfnamefont {T.}~\bibnamefont
  {Matos}}\ and\ \bibinfo {author} {\bibfnamefont {F.~S.}\ \bibnamefont
  {Guzman}},\ }\href {\doibase 10.1088/0264-9381/17/1/102} {\bibfield
  {journal} {\bibinfo  {journal} {Class. Quant. Grav.}\ }\textbf {\bibinfo
  {volume} {17}},\ \bibinfo {pages} {L9} (\bibinfo {year} {2000})},\ \Eprint
  {http://arxiv.org/abs/gr-qc/9810028} {arXiv:gr-qc/9810028} \BibitemShut
  {NoStop}%
\bibitem [{\citenamefont {Hui}\ \emph {et~al.}(2019)\citenamefont {Hui},
  \citenamefont {Kabat}, \citenamefont {Li}, \citenamefont {Santoni},\ and\
  \citenamefont {Wong}}]{Hui:2019aqm}%
  \BibitemOpen
  \bibfield  {author} {\bibinfo {author} {\bibfnamefont {L.}~\bibnamefont
  {Hui}}, \bibinfo {author} {\bibfnamefont {D.}~\bibnamefont {Kabat}}, \bibinfo
  {author} {\bibfnamefont {X.}~\bibnamefont {Li}}, \bibinfo {author}
  {\bibfnamefont {L.}~\bibnamefont {Santoni}}, \ and\ \bibinfo {author}
  {\bibfnamefont {S.~S.~C.}\ \bibnamefont {Wong}},\ }\href {\doibase
  10.1088/1475-7516/2019/06/038} {\bibfield  {journal} {\bibinfo  {journal}
  {JCAP}\ }\textbf {\bibinfo {volume} {06}},\ \bibinfo {pages} {038} (\bibinfo
  {year} {2019})},\ \Eprint {http://arxiv.org/abs/1904.12803} {arXiv:1904.12803
  [gr-qc]} \BibitemShut {NoStop}%
\bibitem [{\citenamefont {Dave}\ and\ \citenamefont
  {Goswami}(2023)}]{Dave:2023wjq}%
  \BibitemOpen
  \bibfield  {author} {\bibinfo {author} {\bibfnamefont {B.}~\bibnamefont
  {Dave}}\ and\ \bibinfo {author} {\bibfnamefont {G.}~\bibnamefont {Goswami}},\
  }\href@noop {} {\  (\bibinfo {year} {2023})},\ \Eprint
  {http://arxiv.org/abs/2304.04463} {arXiv:2304.04463 [astro-ph.CO]}
  \BibitemShut {NoStop}%
\bibitem [{\citenamefont {Pantig}\ and\ \citenamefont
  {\"Ovg\"un}(2023)}]{Pantig:2022sjb}%
  \BibitemOpen
  \bibfield  {author} {\bibinfo {author} {\bibfnamefont {R.~C.}\ \bibnamefont
  {Pantig}}\ and\ \bibinfo {author} {\bibfnamefont {A.}~\bibnamefont
  {\"Ovg\"un}},\ }\href {\doibase 10.1002/prop.202200164} {\bibfield  {journal}
  {\bibinfo  {journal} {Fortsch. Phys.}\ }\textbf {\bibinfo {volume} {71}},\
  \bibinfo {pages} {2200164} (\bibinfo {year} {2023})},\ \Eprint
  {http://arxiv.org/abs/2210.00523} {arXiv:2210.00523 [gr-qc]} \BibitemShut
  {NoStop}%
\bibitem [{\citenamefont {Goodman}(2000)}]{Goodman:2000tg}%
  \BibitemOpen
  \bibfield  {author} {\bibinfo {author} {\bibfnamefont {J.}~\bibnamefont
  {Goodman}},\ }\href {\doibase 10.1016/S1384-1076(00)00015-4} {\bibfield
  {journal} {\bibinfo  {journal} {New Astron.}\ }\textbf {\bibinfo {volume}
  {5}},\ \bibinfo {pages} {103} (\bibinfo {year} {2000})},\ \Eprint
  {http://arxiv.org/abs/astro-ph/0003018} {arXiv:astro-ph/0003018} \BibitemShut
  {NoStop}%
\bibitem [{\citenamefont {Peebles}(2000)}]{Peebles:2000yy}%
  \BibitemOpen
  \bibfield  {author} {\bibinfo {author} {\bibfnamefont {P.~J.~E.}\
  \bibnamefont {Peebles}},\ }\href {\doibase 10.1086/312677} {\bibfield
  {journal} {\bibinfo  {journal} {Astrophys. J. Lett.}\ }\textbf {\bibinfo
  {volume} {534}},\ \bibinfo {pages} {L127} (\bibinfo {year} {2000})},\ \Eprint
  {http://arxiv.org/abs/astro-ph/0002495} {arXiv:astro-ph/0002495} \BibitemShut
  {NoStop}%
\bibitem [{\citenamefont {Arbey}\ \emph {et~al.}(2003)\citenamefont {Arbey},
  \citenamefont {Lesgourgues},\ and\ \citenamefont {Salati}}]{Arbey:2003sj}%
  \BibitemOpen
  \bibfield  {author} {\bibinfo {author} {\bibfnamefont {A.}~\bibnamefont
  {Arbey}}, \bibinfo {author} {\bibfnamefont {J.}~\bibnamefont {Lesgourgues}},
  \ and\ \bibinfo {author} {\bibfnamefont {P.}~\bibnamefont {Salati}},\ }\href
  {\doibase 10.1103/PhysRevD.68.023511} {\bibfield  {journal} {\bibinfo
  {journal} {Phys. Rev. D}\ }\textbf {\bibinfo {volume} {68}},\ \bibinfo
  {pages} {023511} (\bibinfo {year} {2003})},\ \Eprint
  {http://arxiv.org/abs/astro-ph/0301533} {arXiv:astro-ph/0301533} \BibitemShut
  {NoStop}%
\bibitem [{\citenamefont {Boehmer}\ and\ \citenamefont
  {Harko}(2007)}]{Boehmer:2007um}%
  \BibitemOpen
  \bibfield  {author} {\bibinfo {author} {\bibfnamefont {C.~G.}\ \bibnamefont
  {Boehmer}}\ and\ \bibinfo {author} {\bibfnamefont {T.}~\bibnamefont
  {Harko}},\ }\href {\doibase 10.1088/1475-7516/2007/06/025} {\bibfield
  {journal} {\bibinfo  {journal} {JCAP}\ }\textbf {\bibinfo {volume} {06}},\
  \bibinfo {pages} {025} (\bibinfo {year} {2007})},\ \Eprint
  {http://arxiv.org/abs/0705.4158} {arXiv:0705.4158 [astro-ph]} \BibitemShut
  {NoStop}%
\bibitem [{\citenamefont {Lee}\ and\ \citenamefont {Lim}(2010)}]{Lee:2008jp}%
  \BibitemOpen
  \bibfield  {author} {\bibinfo {author} {\bibfnamefont {J.-W.}\ \bibnamefont
  {Lee}}\ and\ \bibinfo {author} {\bibfnamefont {S.}~\bibnamefont {Lim}},\
  }\href {\doibase 10.1088/1475-7516/2010/01/007} {\bibfield  {journal}
  {\bibinfo  {journal} {JCAP}\ }\textbf {\bibinfo {volume} {01}},\ \bibinfo
  {pages} {007} (\bibinfo {year} {2010})},\ \Eprint
  {http://arxiv.org/abs/0812.1342} {arXiv:0812.1342 [astro-ph]} \BibitemShut
  {NoStop}%
\bibitem [{\citenamefont {Harko}(2011)}]{Harko:2011xw}%
  \BibitemOpen
  \bibfield  {author} {\bibinfo {author} {\bibfnamefont {T.}~\bibnamefont
  {Harko}},\ }\href {\doibase 10.1088/1475-7516/2011/05/022} {\bibfield
  {journal} {\bibinfo  {journal} {JCAP}\ }\textbf {\bibinfo {volume} {05}},\
  \bibinfo {pages} {022} (\bibinfo {year} {2011})},\ \Eprint
  {http://arxiv.org/abs/1105.2996} {arXiv:1105.2996 [astro-ph.CO]} \BibitemShut
  {NoStop}%
\bibitem [{\citenamefont {Rindler-Daller}\ and\ \citenamefont
  {Shapiro}(2012)}]{Rindler-Daller:2011afd}%
  \BibitemOpen
  \bibfield  {author} {\bibinfo {author} {\bibfnamefont {T.}~\bibnamefont
  {Rindler-Daller}}\ and\ \bibinfo {author} {\bibfnamefont {P.~R.}\
  \bibnamefont {Shapiro}},\ }\href {\doibase 10.1111/j.1365-2966.2012.20588.x}
  {\bibfield  {journal} {\bibinfo  {journal} {Mon. Not. Roy. Astron. Soc.}\
  }\textbf {\bibinfo {volume} {422}},\ \bibinfo {pages} {135} (\bibinfo {year}
  {2012})},\ \Eprint {http://arxiv.org/abs/1106.1256} {arXiv:1106.1256
  [astro-ph.CO]} \BibitemShut {NoStop}%
\bibitem [{\citenamefont {Su\'arez}\ \emph {et~al.}(2014)\citenamefont
  {Su\'arez}, \citenamefont {Robles},\ and\ \citenamefont
  {Matos}}]{Suarez:2013iw}%
  \BibitemOpen
  \bibfield  {author} {\bibinfo {author} {\bibfnamefont {A.}~\bibnamefont
  {Su\'arez}}, \bibinfo {author} {\bibfnamefont {V.~H.}\ \bibnamefont
  {Robles}}, \ and\ \bibinfo {author} {\bibfnamefont {T.}~\bibnamefont
  {Matos}},\ }\href {\doibase 10.1007/978-3-319-02063-1_9} {\bibfield
  {journal} {\bibinfo  {journal} {Astrophys. Space Sci. Proc.}\ }\textbf
  {\bibinfo {volume} {38}},\ \bibinfo {pages} {107} (\bibinfo {year} {2014})},\
  \Eprint {http://arxiv.org/abs/1302.0903} {arXiv:1302.0903 [astro-ph.CO]}
  \BibitemShut {NoStop}%
\bibitem [{\citenamefont {Ure\~na L\'opez}(2019)}]{Urena-Lopez:2019kud}%
  \BibitemOpen
  \bibfield  {author} {\bibinfo {author} {\bibfnamefont {L.~A.}\ \bibnamefont
  {Ure\~na L\'opez}},\ }\href {\doibase 10.3389/fspas.2019.00047} {\bibfield
  {journal} {\bibinfo  {journal} {Front. Astron. Space Sci.}\ }\textbf
  {\bibinfo {volume} {6}},\ \bibinfo {pages} {47} (\bibinfo {year}
  {2019})}\BibitemShut {NoStop}%
\bibitem [{\citenamefont {Lee}\ and\ \citenamefont {Koh}(1996)}]{Lee:1995af}%
  \BibitemOpen
  \bibfield  {author} {\bibinfo {author} {\bibfnamefont {J.-w.}\ \bibnamefont
  {Lee}}\ and\ \bibinfo {author} {\bibfnamefont {I.-g.}\ \bibnamefont {Koh}},\
  }\href {\doibase 10.1103/PhysRevD.53.2236} {\bibfield  {journal} {\bibinfo
  {journal} {Phys. Rev. D}\ }\textbf {\bibinfo {volume} {53}},\ \bibinfo
  {pages} {2236} (\bibinfo {year} {1996})},\ \Eprint
  {http://arxiv.org/abs/hep-ph/9507385} {arXiv:hep-ph/9507385} \BibitemShut
  {NoStop}%
\bibitem [{\citenamefont {Peccei}\ and\ \citenamefont
  {Quinn}(1977)}]{Peccei:1977hh}%
  \BibitemOpen
  \bibfield  {author} {\bibinfo {author} {\bibfnamefont {R.~D.}\ \bibnamefont
  {Peccei}}\ and\ \bibinfo {author} {\bibfnamefont {H.~R.}\ \bibnamefont
  {Quinn}},\ }\href {\doibase 10.1103/PhysRevLett.38.1440} {\bibfield
  {journal} {\bibinfo  {journal} {Phys. Rev. Lett.}\ }\textbf {\bibinfo
  {volume} {38}},\ \bibinfo {pages} {1440} (\bibinfo {year}
  {1977})}\BibitemShut {NoStop}%
\bibitem [{\citenamefont {Wilczek}(1978)}]{Wilczek:1977pj}%
  \BibitemOpen
  \bibfield  {author} {\bibinfo {author} {\bibfnamefont {F.}~\bibnamefont
  {Wilczek}},\ }\href {\doibase 10.1103/PhysRevLett.40.279} {\bibfield
  {journal} {\bibinfo  {journal} {Phys. Rev. Lett.}\ }\textbf {\bibinfo
  {volume} {40}},\ \bibinfo {pages} {279} (\bibinfo {year} {1978})}\BibitemShut
  {NoStop}%
\bibitem [{\citenamefont {Weinberg}(1978)}]{Weinberg:1977ma}%
  \BibitemOpen
  \bibfield  {author} {\bibinfo {author} {\bibfnamefont {S.}~\bibnamefont
  {Weinberg}},\ }\href {\doibase 10.1103/PhysRevLett.40.223} {\bibfield
  {journal} {\bibinfo  {journal} {Phys. Rev. Lett.}\ }\textbf {\bibinfo
  {volume} {40}},\ \bibinfo {pages} {223} (\bibinfo {year} {1978})}\BibitemShut
  {NoStop}%
\bibitem [{\citenamefont {Khlopov}\ \emph {et~al.}(1985)\citenamefont
  {Khlopov}, \citenamefont {Malomed}, \citenamefont {Zeldovich},\ and\
  \citenamefont {Zeldovich}}]{Khlopov:1985fch}%
  \BibitemOpen
  \bibfield  {author} {\bibinfo {author} {\bibfnamefont {M.~Y.}\ \bibnamefont
  {Khlopov}}, \bibinfo {author} {\bibfnamefont {B.~A.}\ \bibnamefont
  {Malomed}}, \bibinfo {author} {\bibfnamefont {I.~B.}\ \bibnamefont
  {Zeldovich}}, \ and\ \bibinfo {author} {\bibfnamefont {Y.~B.}\ \bibnamefont
  {Zeldovich}},\ }\href {\doibase 10.1093/mnras/215.4.575} {\bibfield
  {journal} {\bibinfo  {journal} {Mon. Not. Roy. Astron. Soc.}\ }\textbf
  {\bibinfo {volume} {215}},\ \bibinfo {pages} {575} (\bibinfo {year}
  {1985})}\BibitemShut {NoStop}%
\bibitem [{\citenamefont {Bhattacharyya}\ \emph {et~al.}(2023)\citenamefont
  {Bhattacharyya}, \citenamefont {Ghosh},\ and\ \citenamefont
  {Pal}}]{Bhattacharyya:2023kbh}%
  \BibitemOpen
  \bibfield  {author} {\bibinfo {author} {\bibfnamefont {A.}~\bibnamefont
  {Bhattacharyya}}, \bibinfo {author} {\bibfnamefont {S.}~\bibnamefont
  {Ghosh}}, \ and\ \bibinfo {author} {\bibfnamefont {S.}~\bibnamefont {Pal}},\
  }\href {\doibase 10.1007/JHEP08(2023)207} {\bibfield  {journal} {\bibinfo
  {journal} {JHEP}\ }\textbf {\bibinfo {volume} {08}},\ \bibinfo {pages} {207}
  (\bibinfo {year} {2023})},\ \Eprint {http://arxiv.org/abs/2305.15473}
  {arXiv:2305.15473 [hep-th]} \BibitemShut {NoStop}%
\bibitem [{\citenamefont {Marsh}(2016)}]{Marsh:2015xka}%
  \BibitemOpen
  \bibfield  {author} {\bibinfo {author} {\bibfnamefont {D.~J.~E.}\
  \bibnamefont {Marsh}},\ }\href {\doibase 10.1016/j.physrep.2016.06.005}
  {\bibfield  {journal} {\bibinfo  {journal} {Phys. Rept.}\ }\textbf {\bibinfo
  {volume} {643}},\ \bibinfo {pages} {1} (\bibinfo {year} {2016})},\ \Eprint
  {http://arxiv.org/abs/1510.07633} {arXiv:1510.07633 [astro-ph.CO]}
  \BibitemShut {NoStop}%
\bibitem [{\citenamefont {Kobayashi}\ \emph {et~al.}(2017)\citenamefont
  {Kobayashi}, \citenamefont {Murgia}, \citenamefont {De~Simone}, \citenamefont
  {Ir\v{s}i\v{c}},\ and\ \citenamefont {Viel}}]{Kobayashi:2017jcf}%
  \BibitemOpen
  \bibfield  {author} {\bibinfo {author} {\bibfnamefont {T.}~\bibnamefont
  {Kobayashi}}, \bibinfo {author} {\bibfnamefont {R.}~\bibnamefont {Murgia}},
  \bibinfo {author} {\bibfnamefont {A.}~\bibnamefont {De~Simone}}, \bibinfo
  {author} {\bibfnamefont {V.}~\bibnamefont {Ir\v{s}i\v{c}}}, \ and\ \bibinfo
  {author} {\bibfnamefont {M.}~\bibnamefont {Viel}},\ }\href {\doibase
  10.1103/PhysRevD.96.123514} {\bibfield  {journal} {\bibinfo  {journal} {Phys.
  Rev. D}\ }\textbf {\bibinfo {volume} {96}},\ \bibinfo {pages} {123514}
  (\bibinfo {year} {2017})},\ \Eprint {http://arxiv.org/abs/1708.00015}
  {arXiv:1708.00015 [astro-ph.CO]} \BibitemShut {NoStop}%
\bibitem [{\citenamefont {Abel}\ \emph {et~al.}(2017)\citenamefont {Abel} \emph
  {et~al.}}]{Abel:2017rtm}%
  \BibitemOpen
  \bibfield  {author} {\bibinfo {author} {\bibfnamefont {C.}~\bibnamefont
  {Abel}} \emph {et~al.},\ }\href {\doibase 10.1103/PhysRevX.7.041034}
  {\bibfield  {journal} {\bibinfo  {journal} {Phys. Rev. X}\ }\textbf {\bibinfo
  {volume} {7}},\ \bibinfo {pages} {041034} (\bibinfo {year} {2017})},\ \Eprint
  {http://arxiv.org/abs/1708.06367} {arXiv:1708.06367 [hep-ph]} \BibitemShut
  {NoStop}%
\bibitem [{\citenamefont {Brito}\ \emph
  {et~al.}(2017{\natexlab{a}})\citenamefont {Brito}, \citenamefont {Ghosh},
  \citenamefont {Barausse}, \citenamefont {Berti}, \citenamefont {Cardoso},
  \citenamefont {Dvorkin}, \citenamefont {Klein},\ and\ \citenamefont
  {Pani}}]{Brito:2017wnc}%
  \BibitemOpen
  \bibfield  {author} {\bibinfo {author} {\bibfnamefont {R.}~\bibnamefont
  {Brito}}, \bibinfo {author} {\bibfnamefont {S.}~\bibnamefont {Ghosh}},
  \bibinfo {author} {\bibfnamefont {E.}~\bibnamefont {Barausse}}, \bibinfo
  {author} {\bibfnamefont {E.}~\bibnamefont {Berti}}, \bibinfo {author}
  {\bibfnamefont {V.}~\bibnamefont {Cardoso}}, \bibinfo {author} {\bibfnamefont
  {I.}~\bibnamefont {Dvorkin}}, \bibinfo {author} {\bibfnamefont
  {A.}~\bibnamefont {Klein}}, \ and\ \bibinfo {author} {\bibfnamefont
  {P.}~\bibnamefont {Pani}},\ }\href {\doibase 10.1103/PhysRevLett.119.131101}
  {\bibfield  {journal} {\bibinfo  {journal} {Phys. Rev. Lett.}\ }\textbf
  {\bibinfo {volume} {119}},\ \bibinfo {pages} {131101} (\bibinfo {year}
  {2017}{\natexlab{a}})},\ \Eprint {http://arxiv.org/abs/1706.05097}
  {arXiv:1706.05097 [gr-qc]} \BibitemShut {NoStop}%
\bibitem [{\citenamefont {Brito}\ \emph
  {et~al.}(2017{\natexlab{b}})\citenamefont {Brito}, \citenamefont {Ghosh},
  \citenamefont {Barausse}, \citenamefont {Berti}, \citenamefont {Cardoso},
  \citenamefont {Dvorkin}, \citenamefont {Klein},\ and\ \citenamefont
  {Pani}}]{Brito:2017zvb}%
  \BibitemOpen
  \bibfield  {author} {\bibinfo {author} {\bibfnamefont {R.}~\bibnamefont
  {Brito}}, \bibinfo {author} {\bibfnamefont {S.}~\bibnamefont {Ghosh}},
  \bibinfo {author} {\bibfnamefont {E.}~\bibnamefont {Barausse}}, \bibinfo
  {author} {\bibfnamefont {E.}~\bibnamefont {Berti}}, \bibinfo {author}
  {\bibfnamefont {V.}~\bibnamefont {Cardoso}}, \bibinfo {author} {\bibfnamefont
  {I.}~\bibnamefont {Dvorkin}}, \bibinfo {author} {\bibfnamefont
  {A.}~\bibnamefont {Klein}}, \ and\ \bibinfo {author} {\bibfnamefont
  {P.}~\bibnamefont {Pani}},\ }\href {\doibase 10.1103/PhysRevD.96.064050}
  {\bibfield  {journal} {\bibinfo  {journal} {Phys. Rev. D}\ }\textbf {\bibinfo
  {volume} {96}},\ \bibinfo {pages} {064050} (\bibinfo {year}
  {2017}{\natexlab{b}})},\ \Eprint {http://arxiv.org/abs/1706.06311}
  {arXiv:1706.06311 [gr-qc]} \BibitemShut {NoStop}%
\bibitem [{\citenamefont {Dave}\ and\ \citenamefont
  {Goswami}(2024)}]{Dave:2023egr}%
  \BibitemOpen
  \bibfield  {author} {\bibinfo {author} {\bibfnamefont {B.}~\bibnamefont
  {Dave}}\ and\ \bibinfo {author} {\bibfnamefont {G.}~\bibnamefont {Goswami}},\
  }\href {\doibase 10.1088/1475-7516/2024/02/044} {\bibfield  {journal}
  {\bibinfo  {journal} {JCAP}\ }\textbf {\bibinfo {volume} {02}},\ \bibinfo
  {pages} {044} (\bibinfo {year} {2024})},\ \Eprint
  {http://arxiv.org/abs/2310.19664} {arXiv:2310.19664 [astro-ph.CO]}
  \BibitemShut {NoStop}%
\bibitem [{\citenamefont {Arbey}\ \emph {et~al.}(2001)\citenamefont {Arbey},
  \citenamefont {Lesgourgues},\ and\ \citenamefont {Salati}}]{Arbey:2001qi}%
  \BibitemOpen
  \bibfield  {author} {\bibinfo {author} {\bibfnamefont {A.}~\bibnamefont
  {Arbey}}, \bibinfo {author} {\bibfnamefont {J.}~\bibnamefont {Lesgourgues}},
  \ and\ \bibinfo {author} {\bibfnamefont {P.}~\bibnamefont {Salati}},\ }\href
  {\doibase 10.1103/PhysRevD.64.123528} {\bibfield  {journal} {\bibinfo
  {journal} {Phys. Rev. D}\ }\textbf {\bibinfo {volume} {64}},\ \bibinfo
  {pages} {123528} (\bibinfo {year} {2001})},\ \Eprint
  {http://arxiv.org/abs/astro-ph/0105564} {arXiv:astro-ph/0105564} \BibitemShut
  {NoStop}%
\bibitem [{\citenamefont {Schive}\ \emph {et~al.}(2014)\citenamefont {Schive},
  \citenamefont {Liao}, \citenamefont {Woo}, \citenamefont {Wong},
  \citenamefont {Chiueh}, \citenamefont {Broadhurst},\ and\ \citenamefont
  {Hwang}}]{Schive:2014hza}%
  \BibitemOpen
  \bibfield  {author} {\bibinfo {author} {\bibfnamefont {H.-Y.}\ \bibnamefont
  {Schive}}, \bibinfo {author} {\bibfnamefont {M.-H.}\ \bibnamefont {Liao}},
  \bibinfo {author} {\bibfnamefont {T.-P.}\ \bibnamefont {Woo}}, \bibinfo
  {author} {\bibfnamefont {S.-K.}\ \bibnamefont {Wong}}, \bibinfo {author}
  {\bibfnamefont {T.}~\bibnamefont {Chiueh}}, \bibinfo {author} {\bibfnamefont
  {T.}~\bibnamefont {Broadhurst}}, \ and\ \bibinfo {author} {\bibfnamefont
  {W.~Y.~P.}\ \bibnamefont {Hwang}},\ }\href {\doibase
  10.1103/PhysRevLett.113.261302} {\bibfield  {journal} {\bibinfo  {journal}
  {Phys. Rev. Lett.}\ }\textbf {\bibinfo {volume} {113}},\ \bibinfo {pages}
  {261302} (\bibinfo {year} {2014})},\ \Eprint {http://arxiv.org/abs/1407.7762}
  {arXiv:1407.7762 [astro-ph.GA]} \BibitemShut {NoStop}%
\bibitem [{\citenamefont {Marsh}\ and\ \citenamefont
  {Pop}(2015)}]{Marsh:2015wka}%
  \BibitemOpen
  \bibfield  {author} {\bibinfo {author} {\bibfnamefont {D.~J.~E.}\
  \bibnamefont {Marsh}}\ and\ \bibinfo {author} {\bibfnamefont {A.-R.}\
  \bibnamefont {Pop}},\ }\href {\doibase 10.1093/mnras/stv1050} {\bibfield
  {journal} {\bibinfo  {journal} {Mon. Not. Roy. Astron. Soc.}\ }\textbf
  {\bibinfo {volume} {451}},\ \bibinfo {pages} {2479} (\bibinfo {year}
  {2015})},\ \Eprint {http://arxiv.org/abs/1502.03456} {arXiv:1502.03456
  [astro-ph.CO]} \BibitemShut {NoStop}%
\bibitem [{\citenamefont {Schwabe}\ \emph {et~al.}(2016)\citenamefont
  {Schwabe}, \citenamefont {Niemeyer},\ and\ \citenamefont
  {Engels}}]{Schwabe:2016rze}%
  \BibitemOpen
  \bibfield  {author} {\bibinfo {author} {\bibfnamefont {B.}~\bibnamefont
  {Schwabe}}, \bibinfo {author} {\bibfnamefont {J.~C.}\ \bibnamefont
  {Niemeyer}}, \ and\ \bibinfo {author} {\bibfnamefont {J.~F.}\ \bibnamefont
  {Engels}},\ }\href {\doibase 10.1103/PhysRevD.94.043513} {\bibfield
  {journal} {\bibinfo  {journal} {Phys. Rev. D}\ }\textbf {\bibinfo {volume}
  {94}},\ \bibinfo {pages} {043513} (\bibinfo {year} {2016})},\ \Eprint
  {http://arxiv.org/abs/1606.05151} {arXiv:1606.05151 [astro-ph.CO]}
  \BibitemShut {NoStop}%
\bibitem [{\citenamefont {Mocz}\ \emph {et~al.}(2017)\citenamefont {Mocz},
  \citenamefont {Vogelsberger}, \citenamefont {Robles}, \citenamefont {Zavala},
  \citenamefont {Boylan-Kolchin}, \citenamefont {Fialkov},\ and\ \citenamefont
  {Hernquist}}]{Mocz:2017wlg}%
  \BibitemOpen
  \bibfield  {author} {\bibinfo {author} {\bibfnamefont {P.}~\bibnamefont
  {Mocz}}, \bibinfo {author} {\bibfnamefont {M.}~\bibnamefont {Vogelsberger}},
  \bibinfo {author} {\bibfnamefont {V.~H.}\ \bibnamefont {Robles}}, \bibinfo
  {author} {\bibfnamefont {J.}~\bibnamefont {Zavala}}, \bibinfo {author}
  {\bibfnamefont {M.}~\bibnamefont {Boylan-Kolchin}}, \bibinfo {author}
  {\bibfnamefont {A.}~\bibnamefont {Fialkov}}, \ and\ \bibinfo {author}
  {\bibfnamefont {L.}~\bibnamefont {Hernquist}},\ }\href {\doibase
  10.1093/mnras/stx1887} {\bibfield  {journal} {\bibinfo  {journal} {Mon. Not.
  Roy. Astron. Soc.}\ }\textbf {\bibinfo {volume} {471}},\ \bibinfo {pages}
  {4559} (\bibinfo {year} {2017})},\ \Eprint {http://arxiv.org/abs/1705.05845}
  {arXiv:1705.05845 [astro-ph.CO]} \BibitemShut {NoStop}%
\bibitem [{\citenamefont {Levkov}\ \emph {et~al.}(2018)\citenamefont {Levkov},
  \citenamefont {Panin},\ and\ \citenamefont {Tkachev}}]{Levkov:2018kau}%
  \BibitemOpen
  \bibfield  {author} {\bibinfo {author} {\bibfnamefont {D.~G.}\ \bibnamefont
  {Levkov}}, \bibinfo {author} {\bibfnamefont {A.~G.}\ \bibnamefont {Panin}}, \
  and\ \bibinfo {author} {\bibfnamefont {I.~I.}\ \bibnamefont {Tkachev}},\
  }\href {\doibase 10.1103/PhysRevLett.121.151301} {\bibfield  {journal}
  {\bibinfo  {journal} {Phys. Rev. Lett.}\ }\textbf {\bibinfo {volume} {121}},\
  \bibinfo {pages} {151301} (\bibinfo {year} {2018})},\ \Eprint
  {http://arxiv.org/abs/1804.05857} {arXiv:1804.05857 [astro-ph.CO]}
  \BibitemShut {NoStop}%
\bibitem [{\citenamefont {{Chavanis}}(2011)}]{2011PhRvD..84d3531C}%
  \BibitemOpen
  \bibfield  {author} {\bibinfo {author} {\bibfnamefont {P.-H.}\ \bibnamefont
  {{Chavanis}}},\ }\href {\doibase 10.1103/PhysRevD.84.043531} {\bibfield
  {journal} {\bibinfo  {journal} {\prd}\ }\textbf {\bibinfo {volume} {84}},\
  \bibinfo {eid} {043531} (\bibinfo {year} {2011})},\ \Eprint
  {http://arxiv.org/abs/1103.2050} {arXiv:1103.2050 [astro-ph.CO]} \BibitemShut
  {NoStop}%
\bibitem [{\citenamefont {{Chavanis}}\ and\ \citenamefont
  {{Delfini}}(2011)}]{2011PhRvD..84d3532C}%
  \BibitemOpen
  \bibfield  {author} {\bibinfo {author} {\bibfnamefont {P.-H.}\ \bibnamefont
  {{Chavanis}}}\ and\ \bibinfo {author} {\bibfnamefont {L.}~\bibnamefont
  {{Delfini}}},\ }\href {\doibase 10.1103/PhysRevD.84.043532} {\bibfield
  {journal} {\bibinfo  {journal} {\prd}\ }\textbf {\bibinfo {volume} {84}},\
  \bibinfo {eid} {043532} (\bibinfo {year} {2011})},\ \Eprint
  {http://arxiv.org/abs/1103.2054} {arXiv:1103.2054 [astro-ph.CO]} \BibitemShut
  {NoStop}%
\bibitem [{\citenamefont {{Schneider}}(2015)}]{2015eaci.book.....S}%
  \BibitemOpen
  \bibfield  {author} {\bibinfo {author} {\bibfnamefont {P.}~\bibnamefont
  {{Schneider}}},\ }\href {\doibase 10.1007/978-3-642-54083-7} {\emph {\bibinfo
  {title} {{Extragalactic Astronomy and Cosmology: An Introduction}}}}\
  (\bibinfo {year} {2015})\BibitemShut {NoStop}%
\bibitem [{\citenamefont {Peebles}(1969)}]{Peebles:1969jm}%
  \BibitemOpen
  \bibfield  {author} {\bibinfo {author} {\bibfnamefont {P.~J.~E.}\
  \bibnamefont {Peebles}},\ }\href {\doibase 10.1086/149876} {\bibfield
  {journal} {\bibinfo  {journal} {Astrophys. J.}\ }\textbf {\bibinfo {volume}
  {155}},\ \bibinfo {pages} {393} (\bibinfo {year} {1969})}\BibitemShut
  {NoStop}%
\bibitem [{\citenamefont {Kain}\ and\ \citenamefont
  {Ling}(2010)}]{Kain:2010rb}%
  \BibitemOpen
  \bibfield  {author} {\bibinfo {author} {\bibfnamefont {B.}~\bibnamefont
  {Kain}}\ and\ \bibinfo {author} {\bibfnamefont {H.~Y.}\ \bibnamefont
  {Ling}},\ }\href {\doibase 10.1103/PhysRevD.82.064042} {\bibfield  {journal}
  {\bibinfo  {journal} {Phys. Rev. D}\ }\textbf {\bibinfo {volume} {82}},\
  \bibinfo {pages} {064042} (\bibinfo {year} {2010})},\ \Eprint
  {http://arxiv.org/abs/1004.4692} {arXiv:1004.4692 [hep-ph]} \BibitemShut
  {NoStop}%
\bibitem [{\citenamefont {Ir\v{s}i\v{c}}\ \emph {et~al.}(2017)\citenamefont
  {Ir\v{s}i\v{c}}, \citenamefont {Viel}, \citenamefont {Haehnelt},
  \citenamefont {Bolton},\ and\ \citenamefont {Becker}}]{Irsic:2017yje}%
  \BibitemOpen
  \bibfield  {author} {\bibinfo {author} {\bibfnamefont {V.}~\bibnamefont
  {Ir\v{s}i\v{c}}}, \bibinfo {author} {\bibfnamefont {M.}~\bibnamefont {Viel}},
  \bibinfo {author} {\bibfnamefont {M.~G.}\ \bibnamefont {Haehnelt}}, \bibinfo
  {author} {\bibfnamefont {J.~S.}\ \bibnamefont {Bolton}}, \ and\ \bibinfo
  {author} {\bibfnamefont {G.~D.}\ \bibnamefont {Becker}},\ }\href {\doibase
  10.1103/PhysRevLett.119.031302} {\bibfield  {journal} {\bibinfo  {journal}
  {Phys. Rev. Lett.}\ }\textbf {\bibinfo {volume} {119}},\ \bibinfo {pages}
  {031302} (\bibinfo {year} {2017})},\ \Eprint
  {http://arxiv.org/abs/1703.04683} {arXiv:1703.04683 [astro-ph.CO]}
  \BibitemShut {NoStop}%
\bibitem [{\citenamefont {Rogers}\ and\ \citenamefont
  {Peiris}(2021)}]{Rogers:2020ltq}%
  \BibitemOpen
  \bibfield  {author} {\bibinfo {author} {\bibfnamefont {K.~K.}\ \bibnamefont
  {Rogers}}\ and\ \bibinfo {author} {\bibfnamefont {H.~V.}\ \bibnamefont
  {Peiris}},\ }\href {\doibase 10.1103/PhysRevLett.126.071302} {\bibfield
  {journal} {\bibinfo  {journal} {Phys. Rev. Lett.}\ }\textbf {\bibinfo
  {volume} {126}},\ \bibinfo {pages} {071302} (\bibinfo {year} {2021})},\
  \Eprint {http://arxiv.org/abs/2007.12705} {arXiv:2007.12705 [astro-ph.CO]}
  \BibitemShut {NoStop}%
\bibitem [{\citenamefont {Bar}\ \emph {et~al.}(2022)\citenamefont {Bar},
  \citenamefont {Blum},\ and\ \citenamefont {Sun}}]{Bar:2021kti}%
  \BibitemOpen
  \bibfield  {author} {\bibinfo {author} {\bibfnamefont {N.}~\bibnamefont
  {Bar}}, \bibinfo {author} {\bibfnamefont {K.}~\bibnamefont {Blum}}, \ and\
  \bibinfo {author} {\bibfnamefont {C.}~\bibnamefont {Sun}},\ }\href {\doibase
  10.1103/PhysRevD.105.083015} {\bibfield  {journal} {\bibinfo  {journal}
  {Phys. Rev. D}\ }\textbf {\bibinfo {volume} {105}},\ \bibinfo {pages}
  {083015} (\bibinfo {year} {2022})},\ \Eprint
  {http://arxiv.org/abs/2111.03070} {arXiv:2111.03070 [hep-ph]} \BibitemShut
  {NoStop}%
\bibitem [{\citenamefont {Kormendy}\ and\ \citenamefont
  {Richstone}(1995)}]{Kormendy:1995er}%
  \BibitemOpen
  \bibfield  {author} {\bibinfo {author} {\bibfnamefont {J.}~\bibnamefont
  {Kormendy}}\ and\ \bibinfo {author} {\bibfnamefont {D.}~\bibnamefont
  {Richstone}},\ }\href {\doibase 10.1146/annurev.aa.33.090195.003053}
  {\bibfield  {journal} {\bibinfo  {journal} {Ann. Rev. Astron. Astrophys.}\
  }\textbf {\bibinfo {volume} {33}},\ \bibinfo {pages} {581} (\bibinfo {year}
  {1995})}\BibitemShut {NoStop}%
\bibitem [{\citenamefont {Ferrarese}\ and\ \citenamefont
  {Ford}(2005)}]{Ferrarese:2004qr}%
  \BibitemOpen
  \bibfield  {author} {\bibinfo {author} {\bibfnamefont {L.}~\bibnamefont
  {Ferrarese}}\ and\ \bibinfo {author} {\bibfnamefont {H.}~\bibnamefont
  {Ford}},\ }\href {\doibase 10.1007/s11214-005-3947-6} {\bibfield  {journal}
  {\bibinfo  {journal} {Space Sci. Rev.}\ }\textbf {\bibinfo {volume} {116}},\
  \bibinfo {pages} {523} (\bibinfo {year} {2005})},\ \Eprint
  {http://arxiv.org/abs/astro-ph/0411247} {arXiv:astro-ph/0411247} \BibitemShut
  {NoStop}%
\bibitem [{\citenamefont {Narayan}(2005)}]{Narayan:2005ie}%
  \BibitemOpen
  \bibfield  {author} {\bibinfo {author} {\bibfnamefont {R.}~\bibnamefont
  {Narayan}},\ }\href {\doibase 10.1088/1367-2630/7/1/199} {\bibfield
  {journal} {\bibinfo  {journal} {New J. Phys.}\ }\textbf {\bibinfo {volume}
  {7}},\ \bibinfo {pages} {199} (\bibinfo {year} {2005})},\ \Eprint
  {http://arxiv.org/abs/gr-qc/0506078} {arXiv:gr-qc/0506078} \BibitemShut
  {NoStop}%
\bibitem [{\citenamefont {Zakharov}(2014)}]{Zakharov:2014lqa}%
  \BibitemOpen
  \bibfield  {author} {\bibinfo {author} {\bibfnamefont {A.~F.}\ \bibnamefont
  {Zakharov}},\ }\href {\doibase 10.1103/PhysRevD.90.062007} {\bibfield
  {journal} {\bibinfo  {journal} {Phys. Rev. D}\ }\textbf {\bibinfo {volume}
  {90}},\ \bibinfo {pages} {062007} (\bibinfo {year} {2014})},\ \Eprint
  {http://arxiv.org/abs/1407.7457} {arXiv:1407.7457 [gr-qc]} \BibitemShut
  {NoStop}%
\bibitem [{\citenamefont {Juraeva}\ \emph {et~al.}(2021)\citenamefont
  {Juraeva}, \citenamefont {Rayimbaev}, \citenamefont {Abdujabbarov},
  \citenamefont {Ahmedov},\ and\ \citenamefont {Palvanov}}]{Juraeva:2021gwb}%
  \BibitemOpen
  \bibfield  {author} {\bibinfo {author} {\bibfnamefont {N.}~\bibnamefont
  {Juraeva}}, \bibinfo {author} {\bibfnamefont {J.}~\bibnamefont {Rayimbaev}},
  \bibinfo {author} {\bibfnamefont {A.}~\bibnamefont {Abdujabbarov}}, \bibinfo
  {author} {\bibfnamefont {B.}~\bibnamefont {Ahmedov}}, \ and\ \bibinfo
  {author} {\bibfnamefont {S.}~\bibnamefont {Palvanov}},\ }\href {\doibase
  10.1140/epjc/s10052-021-08876-5} {\bibfield  {journal} {\bibinfo  {journal}
  {Eur. Phys. J. C}\ }\textbf {\bibinfo {volume} {81}},\ \bibinfo {pages} {70}
  (\bibinfo {year} {2021})}\BibitemShut {NoStop}%
\bibitem [{\citenamefont {Kocherlakota}\ \emph {et~al.}(2021)\citenamefont
  {Kocherlakota} \emph {et~al.}}]{EventHorizonTelescope:2021dqv}%
  \BibitemOpen
  \bibfield  {author} {\bibinfo {author} {\bibfnamefont {P.}~\bibnamefont
  {Kocherlakota}} \emph {et~al.} (\bibinfo {collaboration} {Event Horizon
  Telescope}),\ }\href {\doibase 10.1103/PhysRevD.103.104047} {\bibfield
  {journal} {\bibinfo  {journal} {Phys. Rev. D}\ }\textbf {\bibinfo {volume}
  {103}},\ \bibinfo {pages} {104047} (\bibinfo {year} {2021})},\ \Eprint
  {http://arxiv.org/abs/2105.09343} {arXiv:2105.09343 [gr-qc]} \BibitemShut
  {NoStop}%
\bibitem [{\citenamefont {Akiyama}\ \emph {et~al.}(2022)\citenamefont {Akiyama}
  \emph {et~al.}}]{EventHorizonTelescope:2022xqj}%
  \BibitemOpen
  \bibfield  {author} {\bibinfo {author} {\bibfnamefont {K.}~\bibnamefont
  {Akiyama}} \emph {et~al.} (\bibinfo {collaboration} {Event Horizon
  Telescope}),\ }\href {\doibase 10.3847/2041-8213/ac6756} {\bibfield
  {journal} {\bibinfo  {journal} {Astrophys. J. Lett.}\ }\textbf {\bibinfo
  {volume} {930}},\ \bibinfo {pages} {L17} (\bibinfo {year}
  {2022})}\BibitemShut {NoStop}%
\bibitem [{\citenamefont {Vagnozzi}\ \emph {et~al.}(2022)\citenamefont
  {Vagnozzi} \emph {et~al.}}]{Vagnozzi:2022moj}%
  \BibitemOpen
  \bibfield  {author} {\bibinfo {author} {\bibfnamefont {S.}~\bibnamefont
  {Vagnozzi}} \emph {et~al.},\ }\href@noop {} {\  (\bibinfo {year} {2022})},\
  \Eprint {http://arxiv.org/abs/2205.07787} {arXiv:2205.07787 [gr-qc]}
  \BibitemShut {NoStop}%
\bibitem [{\citenamefont {Abbott}\ \emph {et~al.}(2016)\citenamefont {Abbott}
  \emph {et~al.}}]{LIGOScientific:2016aoc}%
  \BibitemOpen
  \bibfield  {author} {\bibinfo {author} {\bibfnamefont {B.~P.}\ \bibnamefont
  {Abbott}} \emph {et~al.} (\bibinfo {collaboration} {LIGO Scientific,
  Virgo}),\ }\href {\doibase 10.1103/PhysRevLett.116.061102} {\bibfield
  {journal} {\bibinfo  {journal} {Phys. Rev. Lett.}\ }\textbf {\bibinfo
  {volume} {116}},\ \bibinfo {pages} {061102} (\bibinfo {year} {2016})},\
  \Eprint {http://arxiv.org/abs/1602.03837} {arXiv:1602.03837 [gr-qc]}
  \BibitemShut {NoStop}%
\bibitem [{\citenamefont {Eda}\ \emph {et~al.}(2013)\citenamefont {Eda},
  \citenamefont {Itoh}, \citenamefont {Kuroyanagi},\ and\ \citenamefont
  {Silk}}]{Eda:2013gg}%
  \BibitemOpen
  \bibfield  {author} {\bibinfo {author} {\bibfnamefont {K.}~\bibnamefont
  {Eda}}, \bibinfo {author} {\bibfnamefont {Y.}~\bibnamefont {Itoh}}, \bibinfo
  {author} {\bibfnamefont {S.}~\bibnamefont {Kuroyanagi}}, \ and\ \bibinfo
  {author} {\bibfnamefont {J.}~\bibnamefont {Silk}},\ }\href {\doibase
  10.1103/PhysRevLett.110.221101} {\bibfield  {journal} {\bibinfo  {journal}
  {Phys. Rev. Lett.}\ }\textbf {\bibinfo {volume} {110}},\ \bibinfo {pages}
  {221101} (\bibinfo {year} {2013})},\ \Eprint {http://arxiv.org/abs/1301.5971}
  {arXiv:1301.5971 [gr-qc]} \BibitemShut {NoStop}%
\bibitem [{\citenamefont {Lacroix}\ and\ \citenamefont
  {Silk}(2013)}]{Lacroix:2012nz}%
  \BibitemOpen
  \bibfield  {author} {\bibinfo {author} {\bibfnamefont {T.}~\bibnamefont
  {Lacroix}}\ and\ \bibinfo {author} {\bibfnamefont {J.}~\bibnamefont {Silk}},\
  }\href {\doibase 10.1051/0004-6361/201220753} {\bibfield  {journal} {\bibinfo
   {journal} {Astron. Astrophys.}\ }\textbf {\bibinfo {volume} {554}},\
  \bibinfo {pages} {A36} (\bibinfo {year} {2013})},\ \Eprint
  {http://arxiv.org/abs/1211.4861} {arXiv:1211.4861 [astro-ph.GA]} \BibitemShut
  {NoStop}%
\bibitem [{\citenamefont {Kavanagh}\ \emph {et~al.}(2020)\citenamefont
  {Kavanagh}, \citenamefont {Nichols}, \citenamefont {Bertone},\ and\
  \citenamefont {Gaggero}}]{Kavanagh:2020cfn}%
  \BibitemOpen
  \bibfield  {author} {\bibinfo {author} {\bibfnamefont {B.~J.}\ \bibnamefont
  {Kavanagh}}, \bibinfo {author} {\bibfnamefont {D.~A.}\ \bibnamefont
  {Nichols}}, \bibinfo {author} {\bibfnamefont {G.}~\bibnamefont {Bertone}}, \
  and\ \bibinfo {author} {\bibfnamefont {D.}~\bibnamefont {Gaggero}},\ }\href
  {\doibase 10.1103/PhysRevD.102.083006} {\bibfield  {journal} {\bibinfo
  {journal} {Phys. Rev. D}\ }\textbf {\bibinfo {volume} {102}},\ \bibinfo
  {pages} {083006} (\bibinfo {year} {2020})},\ \Eprint
  {http://arxiv.org/abs/2002.12811} {arXiv:2002.12811 [gr-qc]} \BibitemShut
  {NoStop}%
\bibitem [{\citenamefont {G\'omez}\ \emph {et~al.}(2016)\citenamefont
  {G\'omez}, \citenamefont {Arg\"uelles}, \citenamefont {Perlick},
  \citenamefont {Rueda},\ and\ \citenamefont {Ruffini}}]{Gomez:2016iod}%
  \BibitemOpen
  \bibfield  {author} {\bibinfo {author} {\bibfnamefont {L.~G.}\ \bibnamefont
  {G\'omez}}, \bibinfo {author} {\bibfnamefont {C.~R.}\ \bibnamefont
  {Arg\"uelles}}, \bibinfo {author} {\bibfnamefont {V.}~\bibnamefont
  {Perlick}}, \bibinfo {author} {\bibfnamefont {J.~A.}\ \bibnamefont {Rueda}},
  \ and\ \bibinfo {author} {\bibfnamefont {R.}~\bibnamefont {Ruffini}},\ }\href
  {\doibase 10.1103/PhysRevD.94.123004} {\bibfield  {journal} {\bibinfo
  {journal} {Phys. Rev. D}\ }\textbf {\bibinfo {volume} {94}},\ \bibinfo
  {pages} {123004} (\bibinfo {year} {2016})},\ \Eprint
  {http://arxiv.org/abs/1610.03442} {arXiv:1610.03442 [astro-ph.GA]}
  \BibitemShut {NoStop}%
\bibitem [{\citenamefont {Bar}\ \emph {et~al.}(2019)\citenamefont {Bar},
  \citenamefont {Blum}, \citenamefont {Lacroix},\ and\ \citenamefont
  {Panci}}]{Bar:2019pnz}%
  \BibitemOpen
  \bibfield  {author} {\bibinfo {author} {\bibfnamefont {N.}~\bibnamefont
  {Bar}}, \bibinfo {author} {\bibfnamefont {K.}~\bibnamefont {Blum}}, \bibinfo
  {author} {\bibfnamefont {T.}~\bibnamefont {Lacroix}}, \ and\ \bibinfo
  {author} {\bibfnamefont {P.}~\bibnamefont {Panci}},\ }\href {\doibase
  10.1088/1475-7516/2019/07/045} {\bibfield  {journal} {\bibinfo  {journal}
  {JCAP}\ }\textbf {\bibinfo {volume} {07}},\ \bibinfo {pages} {045} (\bibinfo
  {year} {2019})},\ \Eprint {http://arxiv.org/abs/1905.11745} {arXiv:1905.11745
  [astro-ph.CO]} \BibitemShut {NoStop}%
\bibitem [{\citenamefont {Saurabh}\ and\ \citenamefont
  {Jusufi}(2021)}]{Saurabh:2020zqg}%
  \BibitemOpen
  \bibfield  {author} {\bibinfo {author} {\bibfnamefont {K.}~\bibnamefont
  {Saurabh}}\ and\ \bibinfo {author} {\bibfnamefont {K.}~\bibnamefont
  {Jusufi}},\ }\href {\doibase 10.1140/epjc/s10052-021-09280-9} {\bibfield
  {journal} {\bibinfo  {journal} {Eur. Phys. J. C}\ }\textbf {\bibinfo {volume}
  {81}},\ \bibinfo {pages} {490} (\bibinfo {year} {2021})},\ \Eprint
  {http://arxiv.org/abs/2009.10599} {arXiv:2009.10599 [gr-qc]} \BibitemShut
  {NoStop}%
\bibitem [{\citenamefont {Boudon}\ \emph {et~al.}(2023)\citenamefont {Boudon},
  \citenamefont {Brax}, \citenamefont {Valageas},\ and\ \citenamefont
  {Wong}}]{Boudon:2023vzl}%
  \BibitemOpen
  \bibfield  {author} {\bibinfo {author} {\bibfnamefont {A.}~\bibnamefont
  {Boudon}}, \bibinfo {author} {\bibfnamefont {P.}~\bibnamefont {Brax}},
  \bibinfo {author} {\bibfnamefont {P.}~\bibnamefont {Valageas}}, \ and\
  \bibinfo {author} {\bibfnamefont {L.~K.}\ \bibnamefont {Wong}},\ }\href@noop
  {} {\  (\bibinfo {year} {2023})},\ \Eprint {http://arxiv.org/abs/2305.18540}
  {arXiv:2305.18540 [astro-ph.CO]} \BibitemShut {NoStop}%
\bibitem [{\citenamefont {Chakrabarti}\ \emph {et~al.}(2022)\citenamefont
  {Chakrabarti}, \citenamefont {Dave}, \citenamefont {Dutta},\ and\
  \citenamefont {Goswami}}]{Chakrabarti:2022owq}%
  \BibitemOpen
  \bibfield  {author} {\bibinfo {author} {\bibfnamefont {S.}~\bibnamefont
  {Chakrabarti}}, \bibinfo {author} {\bibfnamefont {B.}~\bibnamefont {Dave}},
  \bibinfo {author} {\bibfnamefont {K.}~\bibnamefont {Dutta}}, \ and\ \bibinfo
  {author} {\bibfnamefont {G.}~\bibnamefont {Goswami}},\ }\href {\doibase
  10.1088/1475-7516/2022/09/074} {\bibfield  {journal} {\bibinfo  {journal}
  {JCAP}\ }\textbf {\bibinfo {volume} {09}},\ \bibinfo {pages} {074} (\bibinfo
  {year} {2022})},\ \Eprint {http://arxiv.org/abs/2202.11081} {arXiv:2202.11081
  [astro-ph.CO]} \BibitemShut {NoStop}%
\bibitem [{\citenamefont {Berezhiani}\ \emph {et~al.}(2023)\citenamefont
  {Berezhiani}, \citenamefont {Cintia}, \citenamefont {De~Luca},\ and\
  \citenamefont {Khoury}}]{Berezhiani:2023vlo}%
  \BibitemOpen
  \bibfield  {author} {\bibinfo {author} {\bibfnamefont {L.}~\bibnamefont
  {Berezhiani}}, \bibinfo {author} {\bibfnamefont {G.}~\bibnamefont {Cintia}},
  \bibinfo {author} {\bibfnamefont {V.}~\bibnamefont {De~Luca}}, \ and\
  \bibinfo {author} {\bibfnamefont {J.}~\bibnamefont {Khoury}},\ }\href@noop {}
  {\  (\bibinfo {year} {2023})},\ \Eprint {http://arxiv.org/abs/2311.07672}
  {arXiv:2311.07672 [astro-ph.CO]} \BibitemShut {NoStop}%
\bibitem [{\citenamefont {Rahman}\ \emph {et~al.}(2024)\citenamefont {Rahman},
  \citenamefont {Kumar},\ and\ \citenamefont {Bhattacharyya}}]{Rahman:2023sof}%
  \BibitemOpen
  \bibfield  {author} {\bibinfo {author} {\bibfnamefont {M.}~\bibnamefont
  {Rahman}}, \bibinfo {author} {\bibfnamefont {S.}~\bibnamefont {Kumar}}, \
  and\ \bibinfo {author} {\bibfnamefont {A.}~\bibnamefont {Bhattacharyya}},\
  }\href {\doibase 10.1088/1475-7516/2024/01/035} {\bibfield  {journal}
  {\bibinfo  {journal} {JCAP}\ }\textbf {\bibinfo {volume} {01}},\ \bibinfo
  {pages} {035} (\bibinfo {year} {2024})},\ \Eprint
  {http://arxiv.org/abs/2306.14971} {arXiv:2306.14971 [gr-qc]} \BibitemShut
  {NoStop}%
\bibitem [{\citenamefont {Sadeghian}\ \emph {et~al.}(2013)\citenamefont
  {Sadeghian}, \citenamefont {Ferrer},\ and\ \citenamefont
  {Will}}]{Sadeghian:2013laa}%
  \BibitemOpen
  \bibfield  {author} {\bibinfo {author} {\bibfnamefont {L.}~\bibnamefont
  {Sadeghian}}, \bibinfo {author} {\bibfnamefont {F.}~\bibnamefont {Ferrer}}, \
  and\ \bibinfo {author} {\bibfnamefont {C.~M.}\ \bibnamefont {Will}},\ }\href
  {\doibase 10.1103/PhysRevD.88.063522} {\bibfield  {journal} {\bibinfo
  {journal} {Phys. Rev. D}\ }\textbf {\bibinfo {volume} {88}},\ \bibinfo
  {pages} {063522} (\bibinfo {year} {2013})},\ \Eprint
  {http://arxiv.org/abs/1305.2619} {arXiv:1305.2619 [astro-ph.GA]} \BibitemShut
  {NoStop}%
\bibitem [{\citenamefont {Gondolo}\ and\ \citenamefont
  {Silk}(1999)}]{Gondolo:1999ef}%
  \BibitemOpen
  \bibfield  {author} {\bibinfo {author} {\bibfnamefont {P.}~\bibnamefont
  {Gondolo}}\ and\ \bibinfo {author} {\bibfnamefont {J.}~\bibnamefont {Silk}},\
  }\href {\doibase 10.1103/PhysRevLett.83.1719} {\bibfield  {journal} {\bibinfo
   {journal} {Phys. Rev. Lett.}\ }\textbf {\bibinfo {volume} {83}},\ \bibinfo
  {pages} {1719} (\bibinfo {year} {1999})},\ \Eprint
  {http://arxiv.org/abs/astro-ph/9906391} {arXiv:astro-ph/9906391} \BibitemShut
  {NoStop}%
\bibitem [{\citenamefont {Barranco}\ \emph {et~al.}(2011)\citenamefont
  {Barranco}, \citenamefont {Bernal}, \citenamefont {Degollado}, \citenamefont
  {Diez-Tejedor}, \citenamefont {Megevand}, \citenamefont {Alcubierre},
  \citenamefont {Nunez},\ and\ \citenamefont {Sarbach}}]{Barranco:2011eyw}%
  \BibitemOpen
  \bibfield  {author} {\bibinfo {author} {\bibfnamefont {J.}~\bibnamefont
  {Barranco}}, \bibinfo {author} {\bibfnamefont {A.}~\bibnamefont {Bernal}},
  \bibinfo {author} {\bibfnamefont {J.~C.}\ \bibnamefont {Degollado}}, \bibinfo
  {author} {\bibfnamefont {A.}~\bibnamefont {Diez-Tejedor}}, \bibinfo {author}
  {\bibfnamefont {M.}~\bibnamefont {Megevand}}, \bibinfo {author}
  {\bibfnamefont {M.}~\bibnamefont {Alcubierre}}, \bibinfo {author}
  {\bibfnamefont {D.}~\bibnamefont {Nunez}}, \ and\ \bibinfo {author}
  {\bibfnamefont {O.}~\bibnamefont {Sarbach}},\ }\href {\doibase
  10.1103/PhysRevD.84.083008} {\bibfield  {journal} {\bibinfo  {journal} {Phys.
  Rev. D}\ }\textbf {\bibinfo {volume} {84}},\ \bibinfo {pages} {083008}
  (\bibinfo {year} {2011})},\ \Eprint {http://arxiv.org/abs/1108.0931}
  {arXiv:1108.0931 [gr-qc]} \BibitemShut {NoStop}%
\bibitem [{\citenamefont {Clough}\ \emph {et~al.}(2019)\citenamefont {Clough},
  \citenamefont {Ferreira},\ and\ \citenamefont {Lagos}}]{Clough:2019jpm}%
  \BibitemOpen
  \bibfield  {author} {\bibinfo {author} {\bibfnamefont {K.}~\bibnamefont
  {Clough}}, \bibinfo {author} {\bibfnamefont {P.~G.}\ \bibnamefont
  {Ferreira}}, \ and\ \bibinfo {author} {\bibfnamefont {M.}~\bibnamefont
  {Lagos}},\ }\href {\doibase 10.1103/PhysRevD.100.063014} {\bibfield
  {journal} {\bibinfo  {journal} {Phys. Rev. D}\ }\textbf {\bibinfo {volume}
  {100}},\ \bibinfo {pages} {063014} (\bibinfo {year} {2019})},\ \Eprint
  {http://arxiv.org/abs/1904.12783} {arXiv:1904.12783 [gr-qc]} \BibitemShut
  {NoStop}%
\bibitem [{\citenamefont {Bamber}\ \emph {et~al.}(2021)\citenamefont {Bamber},
  \citenamefont {Clough}, \citenamefont {Ferreira}, \citenamefont {Hui},\ and\
  \citenamefont {Lagos}}]{Bamber:2020bpu}%
  \BibitemOpen
  \bibfield  {author} {\bibinfo {author} {\bibfnamefont {J.}~\bibnamefont
  {Bamber}}, \bibinfo {author} {\bibfnamefont {K.}~\bibnamefont {Clough}},
  \bibinfo {author} {\bibfnamefont {P.~G.}\ \bibnamefont {Ferreira}}, \bibinfo
  {author} {\bibfnamefont {L.}~\bibnamefont {Hui}}, \ and\ \bibinfo {author}
  {\bibfnamefont {M.}~\bibnamefont {Lagos}},\ }\href {\doibase
  10.1103/PhysRevD.103.044059} {\bibfield  {journal} {\bibinfo  {journal}
  {Phys. Rev. D}\ }\textbf {\bibinfo {volume} {103}},\ \bibinfo {pages}
  {044059} (\bibinfo {year} {2021})},\ \Eprint
  {http://arxiv.org/abs/2011.07870} {arXiv:2011.07870 [gr-qc]} \BibitemShut
  {NoStop}%
\bibitem [{\citenamefont {Aguilar-Nieto}\ \emph {et~al.}(2023)\citenamefont
  {Aguilar-Nieto}, \citenamefont {Jaramillo}, \citenamefont {Barranco},
  \citenamefont {Bernal}, \citenamefont {Degollado},\ and\ \citenamefont
  {N\'u\~nez}}]{Aguilar-Nieto:2022jio}%
  \BibitemOpen
  \bibfield  {author} {\bibinfo {author} {\bibfnamefont {A.}~\bibnamefont
  {Aguilar-Nieto}}, \bibinfo {author} {\bibfnamefont {V.}~\bibnamefont
  {Jaramillo}}, \bibinfo {author} {\bibfnamefont {J.}~\bibnamefont {Barranco}},
  \bibinfo {author} {\bibfnamefont {A.}~\bibnamefont {Bernal}}, \bibinfo
  {author} {\bibfnamefont {J.~C.}\ \bibnamefont {Degollado}}, \ and\ \bibinfo
  {author} {\bibfnamefont {D.}~\bibnamefont {N\'u\~nez}},\ }\href {\doibase
  10.1103/PhysRevD.107.044070} {\bibfield  {journal} {\bibinfo  {journal}
  {Phys. Rev. D}\ }\textbf {\bibinfo {volume} {107}},\ \bibinfo {pages}
  {044070} (\bibinfo {year} {2023})},\ \Eprint
  {http://arxiv.org/abs/2211.10456} {arXiv:2211.10456 [gr-qc]} \BibitemShut
  {NoStop}%
\bibitem [{\citenamefont {Brax}\ \emph {et~al.}(2020)\citenamefont {Brax},
  \citenamefont {Cembranos},\ and\ \citenamefont {Valageas}}]{Brax:2019npi}%
  \BibitemOpen
  \bibfield  {author} {\bibinfo {author} {\bibfnamefont {P.}~\bibnamefont
  {Brax}}, \bibinfo {author} {\bibfnamefont {J.~A.~R.}\ \bibnamefont
  {Cembranos}}, \ and\ \bibinfo {author} {\bibfnamefont {P.}~\bibnamefont
  {Valageas}},\ }\href {\doibase 10.1103/PhysRevD.101.023521} {\bibfield
  {journal} {\bibinfo  {journal} {Phys. Rev. D}\ }\textbf {\bibinfo {volume}
  {101}},\ \bibinfo {pages} {023521} (\bibinfo {year} {2020})},\ \Eprint
  {http://arxiv.org/abs/1909.02614} {arXiv:1909.02614 [astro-ph.CO]}
  \BibitemShut {NoStop}%
\bibitem [{\citenamefont {Vicente}\ and\ \citenamefont
  {Cardoso}(2022)}]{Vicente:2022ivh}%
  \BibitemOpen
  \bibfield  {author} {\bibinfo {author} {\bibfnamefont {R.}~\bibnamefont
  {Vicente}}\ and\ \bibinfo {author} {\bibfnamefont {V.}~\bibnamefont
  {Cardoso}},\ }\href {\doibase 10.1103/PhysRevD.105.083008} {\bibfield
  {journal} {\bibinfo  {journal} {Phys. Rev. D}\ }\textbf {\bibinfo {volume}
  {105}},\ \bibinfo {pages} {083008} (\bibinfo {year} {2022})},\ \Eprint
  {http://arxiv.org/abs/2201.08854} {arXiv:2201.08854 [gr-qc]} \BibitemShut
  {NoStop}%
\bibitem [{\citenamefont {Cardoso}\ \emph {et~al.}(2022)\citenamefont
  {Cardoso}, \citenamefont {Ikeda}, \citenamefont {Vicente},\ and\
  \citenamefont {Zilh\~ao}}]{Cardoso:2022nzc}%
  \BibitemOpen
  \bibfield  {author} {\bibinfo {author} {\bibfnamefont {V.}~\bibnamefont
  {Cardoso}}, \bibinfo {author} {\bibfnamefont {T.}~\bibnamefont {Ikeda}},
  \bibinfo {author} {\bibfnamefont {R.}~\bibnamefont {Vicente}}, \ and\
  \bibinfo {author} {\bibfnamefont {M.}~\bibnamefont {Zilh\~ao}},\ }\href
  {\doibase 10.1103/PhysRevD.106.L121302} {\bibfield  {journal} {\bibinfo
  {journal} {Phys. Rev. D}\ }\textbf {\bibinfo {volume} {106}},\ \bibinfo
  {pages} {L121302} (\bibinfo {year} {2022})},\ \Eprint
  {http://arxiv.org/abs/2207.09469} {arXiv:2207.09469 [gr-qc]} \BibitemShut
  {NoStop}%
\bibitem [{\citenamefont {Jacobson}(1999)}]{Jacobson:1999vr}%
  \BibitemOpen
  \bibfield  {author} {\bibinfo {author} {\bibfnamefont {T.}~\bibnamefont
  {Jacobson}},\ }\href {\doibase 10.1103/PhysRevLett.83.2699} {\bibfield
  {journal} {\bibinfo  {journal} {Phys. Rev. Lett.}\ }\textbf {\bibinfo
  {volume} {83}},\ \bibinfo {pages} {2699} (\bibinfo {year} {1999})},\ \Eprint
  {http://arxiv.org/abs/astro-ph/9905303} {arXiv:astro-ph/9905303} \BibitemShut
  {NoStop}%
\bibitem [{\citenamefont {Brihaye}\ and\ \citenamefont
  {Hartmann}(2022)}]{Brihaye:2021phs}%
  \BibitemOpen
  \bibfield  {author} {\bibinfo {author} {\bibfnamefont {Y.}~\bibnamefont
  {Brihaye}}\ and\ \bibinfo {author} {\bibfnamefont {B.}~\bibnamefont
  {Hartmann}},\ }\href {\doibase 10.1088/1361-6382/ac35a9} {\bibfield
  {journal} {\bibinfo  {journal} {Class. Quant. Grav.}\ }\textbf {\bibinfo
  {volume} {39}},\ \bibinfo {pages} {015010} (\bibinfo {year} {2022})},\
  \Eprint {http://arxiv.org/abs/2108.02248} {arXiv:2108.02248 [gr-qc]}
  \BibitemShut {NoStop}%
\bibitem [{\citenamefont {Hong}\ \emph
  {et~al.}(2020{\natexlab{a}})\citenamefont {Hong}, \citenamefont {Suzuki},\
  and\ \citenamefont {Yamada}}]{Hong:2020miv}%
  \BibitemOpen
  \bibfield  {author} {\bibinfo {author} {\bibfnamefont {J.-P.}\ \bibnamefont
  {Hong}}, \bibinfo {author} {\bibfnamefont {M.}~\bibnamefont {Suzuki}}, \ and\
  \bibinfo {author} {\bibfnamefont {M.}~\bibnamefont {Yamada}},\ }\href
  {\doibase 10.1103/PhysRevLett.125.111104} {\bibfield  {journal} {\bibinfo
  {journal} {Phys. Rev. Lett.}\ }\textbf {\bibinfo {volume} {125}},\ \bibinfo
  {pages} {111104} (\bibinfo {year} {2020}{\natexlab{a}})},\ \Eprint
  {http://arxiv.org/abs/2004.03148} {arXiv:2004.03148 [gr-qc]} \BibitemShut
  {NoStop}%
\bibitem [{\citenamefont {Unruh}(1976)}]{Unruh:1976fm}%
  \BibitemOpen
  \bibfield  {author} {\bibinfo {author} {\bibfnamefont {W.~G.}\ \bibnamefont
  {Unruh}},\ }\href {\doibase 10.1103/PhysRevD.14.3251} {\bibfield  {journal}
  {\bibinfo  {journal} {Phys. Rev. D}\ }\textbf {\bibinfo {volume} {14}},\
  \bibinfo {pages} {3251} (\bibinfo {year} {1976})}\BibitemShut {NoStop}%
\bibitem [{\citenamefont {Detweiler}(1980)}]{Detweiler:1980uk}%
  \BibitemOpen
  \bibfield  {author} {\bibinfo {author} {\bibfnamefont {S.~L.}\ \bibnamefont
  {Detweiler}},\ }\href {\doibase 10.1103/PhysRevD.22.2323} {\bibfield
  {journal} {\bibinfo  {journal} {Phys. Rev. D}\ }\textbf {\bibinfo {volume}
  {22}},\ \bibinfo {pages} {2323} (\bibinfo {year} {1980})}\BibitemShut
  {NoStop}%
\bibitem [{\citenamefont {Benone}\ \emph {et~al.}(2014)\citenamefont {Benone},
  \citenamefont {de~Oliveira}, \citenamefont {Dolan},\ and\ \citenamefont
  {Crispino}}]{Benone:2014qaa}%
  \BibitemOpen
  \bibfield  {author} {\bibinfo {author} {\bibfnamefont {C.~L.}\ \bibnamefont
  {Benone}}, \bibinfo {author} {\bibfnamefont {E.~S.}\ \bibnamefont
  {de~Oliveira}}, \bibinfo {author} {\bibfnamefont {S.~R.}\ \bibnamefont
  {Dolan}}, \ and\ \bibinfo {author} {\bibfnamefont {L.~C.~B.}\ \bibnamefont
  {Crispino}},\ }\href {\doibase 10.1103/PhysRevD.89.104053} {\bibfield
  {journal} {\bibinfo  {journal} {Phys. Rev. D}\ }\textbf {\bibinfo {volume}
  {89}},\ \bibinfo {pages} {104053} (\bibinfo {year} {2014})},\ \Eprint
  {http://arxiv.org/abs/1404.0687} {arXiv:1404.0687 [gr-qc]} \BibitemShut
  {NoStop}%
\bibitem [{\citenamefont {Hong}\ \emph
  {et~al.}(2020{\natexlab{b}})\citenamefont {Hong}, \citenamefont {Suzuki},\
  and\ \citenamefont {Yamada}}]{Hong:2019mcj}%
  \BibitemOpen
  \bibfield  {author} {\bibinfo {author} {\bibfnamefont {J.-P.}\ \bibnamefont
  {Hong}}, \bibinfo {author} {\bibfnamefont {M.}~\bibnamefont {Suzuki}}, \ and\
  \bibinfo {author} {\bibfnamefont {M.}~\bibnamefont {Yamada}},\ }\href
  {\doibase 10.1016/j.physletb.2020.135324} {\bibfield  {journal} {\bibinfo
  {journal} {Phys. Lett. B}\ }\textbf {\bibinfo {volume} {803}},\ \bibinfo
  {pages} {135324} (\bibinfo {year} {2020}{\natexlab{b}})},\ \Eprint
  {http://arxiv.org/abs/1907.04982} {arXiv:1907.04982 [gr-qc]} \BibitemShut
  {NoStop}%
\bibitem [{\citenamefont {\"Ovg\"un}\ \emph {et~al.}(2024)\citenamefont
  {\"Ovg\"un}, \citenamefont {Sese},\ and\ \citenamefont
  {Pantig}}]{Ovgun:2023wmc}%
  \BibitemOpen
  \bibfield  {author} {\bibinfo {author} {\bibfnamefont {A.}~\bibnamefont
  {\"Ovg\"un}}, \bibinfo {author} {\bibfnamefont {L.~J.~F.}\ \bibnamefont
  {Sese}}, \ and\ \bibinfo {author} {\bibfnamefont {R.~C.}\ \bibnamefont
  {Pantig}},\ }\href {\doibase 10.1002/andp.202300390} {\bibfield  {journal}
  {\bibinfo  {journal} {Annalen Phys.}\ }\textbf {\bibinfo {volume} {536}},\
  \bibinfo {pages} {2300390} (\bibinfo {year} {2024})},\ \Eprint
  {http://arxiv.org/abs/2309.07442} {arXiv:2309.07442 [gr-qc]} \BibitemShut
  {NoStop}%
\bibitem [{\citenamefont {Babichev}\ \emph {et~al.}(2011)\citenamefont
  {Babichev}, \citenamefont {Chernov}, \citenamefont {Dokuchaev},\ and\
  \citenamefont {Eroshenko}}]{Babichev:2008jb}%
  \BibitemOpen
  \bibfield  {author} {\bibinfo {author} {\bibfnamefont {E.}~\bibnamefont
  {Babichev}}, \bibinfo {author} {\bibfnamefont {S.}~\bibnamefont {Chernov}},
  \bibinfo {author} {\bibfnamefont {V.}~\bibnamefont {Dokuchaev}}, \ and\
  \bibinfo {author} {\bibfnamefont {Y.}~\bibnamefont {Eroshenko}},\ }\href
  {\doibase 10.1134/S1063776111040157} {\bibfield  {journal} {\bibinfo
  {journal} {J. Exp. Theor. Phys.}\ }\textbf {\bibinfo {volume} {112}},\
  \bibinfo {pages} {784} (\bibinfo {year} {2011})},\ \Eprint
  {http://arxiv.org/abs/0806.0916} {arXiv:0806.0916 [gr-qc]} \BibitemShut
  {NoStop}%
\bibitem [{\citenamefont {Reissner}(1916)}]{reissner1916eigengravitation}%
  \BibitemOpen
  \bibfield  {author} {\bibinfo {author} {\bibfnamefont {H.}~\bibnamefont
  {Reissner}},\ }\href@noop {} {\bibfield  {journal} {\bibinfo  {journal}
  {Annalen der Physik}\ }\textbf {\bibinfo {volume} {355}},\ \bibinfo {pages}
  {106} (\bibinfo {year} {1916})}\BibitemShut {NoStop}%
\bibitem [{\citenamefont {Ravanal}\ \emph {et~al.}(2023)\citenamefont
  {Ravanal}, \citenamefont {G\'omez},\ and\ \citenamefont
  {Cruz}}]{Ravanal:2023ytp}%
  \BibitemOpen
  \bibfield  {author} {\bibinfo {author} {\bibfnamefont {Y.}~\bibnamefont
  {Ravanal}}, \bibinfo {author} {\bibfnamefont {G.}~\bibnamefont {G\'omez}}, \
  and\ \bibinfo {author} {\bibfnamefont {N.}~\bibnamefont {Cruz}},\ }\href
  {\doibase 10.1103/PhysRevD.108.083004} {\bibfield  {journal} {\bibinfo
  {journal} {Phys. Rev. D}\ }\textbf {\bibinfo {volume} {108}},\ \bibinfo
  {pages} {083004} (\bibinfo {year} {2023})},\ \Eprint
  {http://arxiv.org/abs/2306.10204} {arXiv:2306.10204 [astro-ph.CO]}
  \BibitemShut {NoStop}%
\bibitem [{\citenamefont {Beheshti}\ and\ \citenamefont
  {Gasperin}(2016)}]{Beheshti:2015bak}%
  \BibitemOpen
  \bibfield  {author} {\bibinfo {author} {\bibfnamefont {S.}~\bibnamefont
  {Beheshti}}\ and\ \bibinfo {author} {\bibfnamefont {E.}~\bibnamefont
  {Gasperin}},\ }\href {\doibase 10.1103/PhysRevD.94.024015} {\bibfield
  {journal} {\bibinfo  {journal} {Phys. Rev. D}\ }\textbf {\bibinfo {volume}
  {94}},\ \bibinfo {pages} {024015} (\bibinfo {year} {2016})},\ \Eprint
  {http://arxiv.org/abs/1512.08707} {arXiv:1512.08707 [gr-qc]} \BibitemShut
  {NoStop}%
\bibitem [{\citenamefont {Hod}(2013)}]{Hod:2013mgr}%
  \BibitemOpen
  \bibfield  {author} {\bibinfo {author} {\bibfnamefont {S.}~\bibnamefont
  {Hod}},\ }\href {\doibase 10.1103/PhysRevD.88.087502} {\bibfield  {journal}
  {\bibinfo  {journal} {Phys. Rev. D}\ }\textbf {\bibinfo {volume} {88}},\
  \bibinfo {pages} {087502} (\bibinfo {year} {2013})},\ \Eprint
  {http://arxiv.org/abs/1707.05680} {arXiv:1707.05680 [gr-qc]} \BibitemShut
  {NoStop}%
\bibitem [{\citenamefont {Feng}\ \emph {et~al.}(2022)\citenamefont {Feng},
  \citenamefont {Parisi}, \citenamefont {Chen},\ and\ \citenamefont
  {Lin}}]{Feng:2021qkj}%
  \BibitemOpen
  \bibfield  {author} {\bibinfo {author} {\bibfnamefont {W.-X.}\ \bibnamefont
  {Feng}}, \bibinfo {author} {\bibfnamefont {A.}~\bibnamefont {Parisi}},
  \bibinfo {author} {\bibfnamefont {C.-S.}\ \bibnamefont {Chen}}, \ and\
  \bibinfo {author} {\bibfnamefont {F.-L.}\ \bibnamefont {Lin}},\ }\href
  {\doibase 10.1088/1475-7516/2022/08/032} {\bibfield  {journal} {\bibinfo
  {journal} {JCAP}\ }\textbf {\bibinfo {volume} {08}},\ \bibinfo {pages} {032}
  (\bibinfo {year} {2022})},\ \Eprint {http://arxiv.org/abs/2112.05160}
  {arXiv:2112.05160 [astro-ph.HE]} \BibitemShut {NoStop}%
\bibitem [{\citenamefont {G\'omez}\ and\ \citenamefont
  {Valageas}(2024)}]{Gomez:2024ack}%
  \BibitemOpen
  \bibfield  {author} {\bibinfo {author} {\bibfnamefont {G.}~\bibnamefont
  {G\'omez}}\ and\ \bibinfo {author} {\bibfnamefont {P.}~\bibnamefont
  {Valageas}},\ }\href {\doibase 10.1103/PhysRevD.109.103038} {\bibfield
  {journal} {\bibinfo  {journal} {Phys. Rev. D}\ }\textbf {\bibinfo {volume}
  {109}},\ \bibinfo {pages} {103038} (\bibinfo {year} {2024})},\ \Eprint
  {http://arxiv.org/abs/2403.08988} {arXiv:2403.08988 [astro-ph.CO]}
  \BibitemShut {NoStop}%
\bibitem [{\citenamefont {Bucciotti}\ and\ \citenamefont
  {Trincherini}(2023)}]{Bucciotti:2023bvw}%
  \BibitemOpen
  \bibfield  {author} {\bibinfo {author} {\bibfnamefont {B.}~\bibnamefont
  {Bucciotti}}\ and\ \bibinfo {author} {\bibfnamefont {E.}~\bibnamefont
  {Trincherini}},\ }\href@noop {} {\  (\bibinfo {year} {2023})},\ \Eprint
  {http://arxiv.org/abs/2309.02482} {arXiv:2309.02482 [hep-th]} \BibitemShut
  {NoStop}%
\bibitem [{\citenamefont {De~Luca}\ and\ \citenamefont
  {Khoury}(2023)}]{DeLuca:2023laa}%
  \BibitemOpen
  \bibfield  {author} {\bibinfo {author} {\bibfnamefont {V.}~\bibnamefont
  {De~Luca}}\ and\ \bibinfo {author} {\bibfnamefont {J.}~\bibnamefont
  {Khoury}},\ }\href {\doibase 10.1088/1475-7516/2023/04/048} {\bibfield
  {journal} {\bibinfo  {journal} {JCAP}\ }\textbf {\bibinfo {volume} {04}},\
  \bibinfo {pages} {048} (\bibinfo {year} {2023})},\ \Eprint
  {http://arxiv.org/abs/2302.10286} {arXiv:2302.10286 [astro-ph.CO]}
  \BibitemShut {NoStop}%
\bibitem [{\citenamefont {Brax}\ \emph {et~al.}(2019)\citenamefont {Brax},
  \citenamefont {Cembranos},\ and\ \citenamefont {Valageas}}]{Brax:2019fzb}%
  \BibitemOpen
  \bibfield  {author} {\bibinfo {author} {\bibfnamefont {P.}~\bibnamefont
  {Brax}}, \bibinfo {author} {\bibfnamefont {J.~A.~R.}\ \bibnamefont
  {Cembranos}}, \ and\ \bibinfo {author} {\bibfnamefont {P.}~\bibnamefont
  {Valageas}},\ }\href {\doibase 10.1103/PhysRevD.100.023526} {\bibfield
  {journal} {\bibinfo  {journal} {Phys. Rev. D}\ }\textbf {\bibinfo {volume}
  {100}},\ \bibinfo {pages} {023526} (\bibinfo {year} {2019})},\ \Eprint
  {http://arxiv.org/abs/1906.00730} {arXiv:1906.00730 [astro-ph.CO]}
  \BibitemShut {NoStop}%
\bibitem [{\citenamefont {Akiyama}\ \emph {et~al.}(2021)\citenamefont {Akiyama}
  \emph {et~al.}}]{EventHorizonTelescope:2021srq}%
  \BibitemOpen
  \bibfield  {author} {\bibinfo {author} {\bibfnamefont {K.}~\bibnamefont
  {Akiyama}} \emph {et~al.} (\bibinfo {collaboration} {Event Horizon
  Telescope}),\ }\href {\doibase 10.3847/2041-8213/abe4de} {\bibfield
  {journal} {\bibinfo  {journal} {Astrophys. J. Lett.}\ }\textbf {\bibinfo
  {volume} {910}},\ \bibinfo {pages} {L13} (\bibinfo {year} {2021})},\ \Eprint
  {http://arxiv.org/abs/2105.01173} {arXiv:2105.01173 [astro-ph.HE]}
  \BibitemShut {NoStop}%
\bibitem [{\citenamefont {Salehian}\ \emph {et~al.}(2021)\citenamefont
  {Salehian}, \citenamefont {Zhang}, \citenamefont {Amin}, \citenamefont
  {Kaiser},\ and\ \citenamefont {Namjoo}}]{Salehian:2021khb}%
  \BibitemOpen
  \bibfield  {author} {\bibinfo {author} {\bibfnamefont {B.}~\bibnamefont
  {Salehian}}, \bibinfo {author} {\bibfnamefont {H.-Y.}\ \bibnamefont {Zhang}},
  \bibinfo {author} {\bibfnamefont {M.~A.}\ \bibnamefont {Amin}}, \bibinfo
  {author} {\bibfnamefont {D.~I.}\ \bibnamefont {Kaiser}}, \ and\ \bibinfo
  {author} {\bibfnamefont {M.~H.}\ \bibnamefont {Namjoo}},\ }\href {\doibase
  10.1007/JHEP09(2021)050} {\bibfield  {journal} {\bibinfo  {journal} {JHEP}\
  }\textbf {\bibinfo {volume} {09}},\ \bibinfo {pages} {050} (\bibinfo {year}
  {2021})},\ \Eprint {http://arxiv.org/abs/2104.10128} {arXiv:2104.10128
  [astro-ph.CO]} \BibitemShut {NoStop}%
\bibitem [{\citenamefont {Madelung}(1927)}]{madelung1927quantum}%
  \BibitemOpen
  \bibfield  {author} {\bibinfo {author} {\bibfnamefont {E.}~\bibnamefont
  {Madelung}},\ }\href@noop {} {\bibfield  {journal} {\bibinfo  {journal} {z.
  Phys}\ }\textbf {\bibinfo {volume} {40}},\ \bibinfo {pages} {322} (\bibinfo
  {year} {1927})}\BibitemShut {NoStop}%
\bibitem [{\citenamefont {Shapiro}\ and\ \citenamefont
  {Teukolsky}(1983)}]{Shapiro:1983du}%
  \BibitemOpen
  \bibfield  {author} {\bibinfo {author} {\bibfnamefont {S.~L.}\ \bibnamefont
  {Shapiro}}\ and\ \bibinfo {author} {\bibfnamefont {S.~A.}\ \bibnamefont
  {Teukolsky}},\ }\href@noop {} {\emph {\bibinfo {title} {{Black holes, white
  dwarfs, and neutron stars: The physics of compact objects}}}}\ (\bibinfo
  {year} {1983})\BibitemShut {NoStop}%
\bibitem [{\citenamefont {TREVES}\ \emph {et~al.}(1988)\citenamefont {TREVES},
  \citenamefont {MARASCHI},\ and\ \citenamefont
  {ABRAMOWICZ}}]{b61f022b-8347-33e4-bf2e-c3d95dd12274}%
  \BibitemOpen
  \bibfield  {author} {\bibinfo {author} {\bibfnamefont {A.}~\bibnamefont
  {TREVES}}, \bibinfo {author} {\bibfnamefont {L.}~\bibnamefont {MARASCHI}}, \
  and\ \bibinfo {author} {\bibfnamefont {M.}~\bibnamefont {ABRAMOWICZ}},\
  }\href {http://www.jstor.org/stable/40679119} {\bibfield  {journal} {\bibinfo
   {journal} {Publications of the Astronomical Society of the Pacific}\
  }\textbf {\bibinfo {volume} {100}},\ \bibinfo {pages} {427} (\bibinfo {year}
  {1988})}\BibitemShut {NoStop}%
\bibitem [{\citenamefont {{Abramowicz}}\ \emph {et~al.}(2010)\citenamefont
  {{Abramowicz}}, \citenamefont {{Jaroszy{\'n}ski}}, \citenamefont {{Kato}},
  \citenamefont {{Lasota}}, \citenamefont {{R{\'o}{\.z}a{\'n}ska}},\ and\
  \citenamefont {{S{\k{a}}dowski}}}]{2010A&A...521A..15A}%
  \BibitemOpen
  \bibfield  {author} {\bibinfo {author} {\bibfnamefont {M.~A.}\ \bibnamefont
  {{Abramowicz}}}, \bibinfo {author} {\bibfnamefont {M.}~\bibnamefont
  {{Jaroszy{\'n}ski}}}, \bibinfo {author} {\bibfnamefont {S.}~\bibnamefont
  {{Kato}}}, \bibinfo {author} {\bibfnamefont {J.~P.}\ \bibnamefont
  {{Lasota}}}, \bibinfo {author} {\bibfnamefont {A.}~\bibnamefont
  {{R{\'o}{\.z}a{\'n}ska}}}, \ and\ \bibinfo {author} {\bibfnamefont
  {A.}~\bibnamefont {{S{\k{a}}dowski}}},\ }\href {\doibase
  10.1051/0004-6361/201014467} {\bibfield  {journal} {\bibinfo  {journal}
  {\aap}\ }\textbf {\bibinfo {volume} {521}},\ \bibinfo {eid} {A15} (\bibinfo
  {year} {2010})},\ \Eprint {http://arxiv.org/abs/1003.3887} {arXiv:1003.3887
  [astro-ph.HE]} \BibitemShut {NoStop}%
\bibitem [{\citenamefont {Pugliese}\ \emph {et~al.}(2011)\citenamefont
  {Pugliese}, \citenamefont {Quevedo},\ and\ \citenamefont
  {Ruffini}}]{Pugliese:2010ps}%
  \BibitemOpen
  \bibfield  {author} {\bibinfo {author} {\bibfnamefont {D.}~\bibnamefont
  {Pugliese}}, \bibinfo {author} {\bibfnamefont {H.}~\bibnamefont {Quevedo}}, \
  and\ \bibinfo {author} {\bibfnamefont {R.}~\bibnamefont {Ruffini}},\ }\href
  {\doibase 10.1103/PhysRevD.83.024021} {\bibfield  {journal} {\bibinfo
  {journal} {Phys. Rev. D}\ }\textbf {\bibinfo {volume} {83}},\ \bibinfo
  {pages} {024021} (\bibinfo {year} {2011})},\ \Eprint
  {http://arxiv.org/abs/1012.5411} {arXiv:1012.5411 [astro-ph.HE]} \BibitemShut
  {NoStop}%
\bibitem [{\citenamefont {Claudel}\ \emph {et~al.}(2001)\citenamefont
  {Claudel}, \citenamefont {Virbhadra},\ and\ \citenamefont
  {Ellis}}]{Claudel:2000yi}%
  \BibitemOpen
  \bibfield  {author} {\bibinfo {author} {\bibfnamefont {C.-M.}\ \bibnamefont
  {Claudel}}, \bibinfo {author} {\bibfnamefont {K.~S.}\ \bibnamefont
  {Virbhadra}}, \ and\ \bibinfo {author} {\bibfnamefont {G.~F.~R.}\
  \bibnamefont {Ellis}},\ }\href {\doibase 10.1063/1.1308507} {\bibfield
  {journal} {\bibinfo  {journal} {J. Math. Phys.}\ }\textbf {\bibinfo {volume}
  {42}},\ \bibinfo {pages} {818} (\bibinfo {year} {2001})},\ \Eprint
  {http://arxiv.org/abs/gr-qc/0005050} {arXiv:gr-qc/0005050} \BibitemShut
  {NoStop}%
\bibitem [{\citenamefont {G\'omez}\ and\ \citenamefont
  {Rodr\'\i{}guez}(2023)}]{Gomez:2023wei}%
  \BibitemOpen
  \bibfield  {author} {\bibinfo {author} {\bibfnamefont {G.}~\bibnamefont
  {G\'omez}}\ and\ \bibinfo {author} {\bibfnamefont {J.~F.}\ \bibnamefont
  {Rodr\'\i{}guez}},\ }\href {\doibase 10.1103/PhysRevD.108.024069} {\bibfield
  {journal} {\bibinfo  {journal} {Phys. Rev. D}\ }\textbf {\bibinfo {volume}
  {108}},\ \bibinfo {pages} {024069} (\bibinfo {year} {2023})},\ \Eprint
  {http://arxiv.org/abs/2301.05222} {arXiv:2301.05222 [gr-qc]} \BibitemShut
  {NoStop}%
\bibitem [{\citenamefont {Kovacic}\ and\ \citenamefont
  {Brennan}(2011)}]{kovacic2011duffing}%
  \BibitemOpen
  \bibfield  {author} {\bibinfo {author} {\bibfnamefont {I.}~\bibnamefont
  {Kovacic}}\ and\ \bibinfo {author} {\bibfnamefont {M.~J.}\ \bibnamefont
  {Brennan}},\ }\href@noop {} {\emph {\bibinfo {title} {The Duffing equation:
  nonlinear oscillators and their behaviour}}}\ (\bibinfo  {publisher} {John
  Wiley \& Sons},\ \bibinfo {year} {2011})\BibitemShut {NoStop}%
\bibitem [{\citenamefont {Frasca}(2011)}]{Frasca:2009bc}%
  \BibitemOpen
  \bibfield  {author} {\bibinfo {author} {\bibfnamefont {M.}~\bibnamefont
  {Frasca}},\ }\href {\doibase 10.1142/S1402925111001441} {\bibfield  {journal}
  {\bibinfo  {journal} {J. Nonlin. Math. Phys.}\ }\textbf {\bibinfo {volume}
  {18}},\ \bibinfo {pages} {291} (\bibinfo {year} {2011})},\ \Eprint
  {http://arxiv.org/abs/0907.4053} {arXiv:0907.4053 [math-ph]} \BibitemShut
  {NoStop}%
\bibitem [{\citenamefont {Fiziev}(2006{\natexlab{a}})}]{Fiziev:2005ki}%
  \BibitemOpen
  \bibfield  {author} {\bibinfo {author} {\bibfnamefont {P.~P.}\ \bibnamefont
  {Fiziev}},\ }\href {\doibase 10.1088/0264-9381/23/7/015} {\bibfield
  {journal} {\bibinfo  {journal} {Class. Quant. Grav.}\ }\textbf {\bibinfo
  {volume} {23}},\ \bibinfo {pages} {2447} (\bibinfo {year}
  {2006}{\natexlab{a}})},\ \Eprint {http://arxiv.org/abs/gr-qc/0509123}
  {arXiv:gr-qc/0509123} \BibitemShut {NoStop}%
\bibitem [{\citenamefont {Fiziev}(2006{\natexlab{b}})}]{Fiziev:2006tx}%
  \BibitemOpen
  \bibfield  {author} {\bibinfo {author} {\bibfnamefont {P.~P.}\ \bibnamefont
  {Fiziev}},\ }\href@noop {} {\  (\bibinfo {year} {2006}{\natexlab{b}})},\
  \Eprint {http://arxiv.org/abs/gr-qc/0603003} {arXiv:gr-qc/0603003}
  \BibitemShut {NoStop}%
\bibitem [{\citenamefont {Bezerra}\ \emph {et~al.}(2014)\citenamefont
  {Bezerra}, \citenamefont {Vieira},\ and\ \citenamefont
  {Costa}}]{Bezerra:2013iha}%
  \BibitemOpen
  \bibfield  {author} {\bibinfo {author} {\bibfnamefont {V.~B.}\ \bibnamefont
  {Bezerra}}, \bibinfo {author} {\bibfnamefont {H.~S.}\ \bibnamefont {Vieira}},
  \ and\ \bibinfo {author} {\bibfnamefont {A.~A.}\ \bibnamefont {Costa}},\
  }\href {\doibase 10.1088/0264-9381/31/4/045003} {\bibfield  {journal}
  {\bibinfo  {journal} {Class. Quant. Grav.}\ }\textbf {\bibinfo {volume}
  {31}},\ \bibinfo {pages} {045003} (\bibinfo {year} {2014})},\ \Eprint
  {http://arxiv.org/abs/1312.4823} {arXiv:1312.4823 [gr-qc]} \BibitemShut
  {NoStop}%
\bibitem [{\citenamefont {Vieira}\ \emph {et~al.}(2014)\citenamefont {Vieira},
  \citenamefont {Bezerra},\ and\ \citenamefont {Muniz}}]{Vieira:2014waa}%
  \BibitemOpen
  \bibfield  {author} {\bibinfo {author} {\bibfnamefont {H.~S.}\ \bibnamefont
  {Vieira}}, \bibinfo {author} {\bibfnamefont {V.~B.}\ \bibnamefont {Bezerra}},
  \ and\ \bibinfo {author} {\bibfnamefont {C.~R.}\ \bibnamefont {Muniz}},\
  }\href {\doibase 10.1016/j.aop.2014.07.011} {\bibfield  {journal} {\bibinfo
  {journal} {Annals Phys.}\ }\textbf {\bibinfo {volume} {350}},\ \bibinfo
  {pages} {14} (\bibinfo {year} {2014})},\ \Eprint
  {http://arxiv.org/abs/1401.5397} {arXiv:1401.5397 [gr-qc]} \BibitemShut
  {NoStop}%
\bibitem [{\citenamefont {Konoplya}\ and\ \citenamefont
  {Zhidenko}(2006)}]{Konoplya:2006br}%
  \BibitemOpen
  \bibfield  {author} {\bibinfo {author} {\bibfnamefont {R.~A.}\ \bibnamefont
  {Konoplya}}\ and\ \bibinfo {author} {\bibfnamefont {A.}~\bibnamefont
  {Zhidenko}},\ }\href {\doibase 10.1103/PhysRevD.73.124040} {\bibfield
  {journal} {\bibinfo  {journal} {Phys. Rev. D}\ }\textbf {\bibinfo {volume}
  {73}},\ \bibinfo {pages} {124040} (\bibinfo {year} {2006})},\ \Eprint
  {http://arxiv.org/abs/gr-qc/0605013} {arXiv:gr-qc/0605013} \BibitemShut
  {NoStop}%
\bibitem [{\citenamefont {Barranco}\ \emph {et~al.}(2012)\citenamefont
  {Barranco}, \citenamefont {Bernal}, \citenamefont {Degollado}, \citenamefont
  {Diez-Tejedor}, \citenamefont {Megevand}, \citenamefont {Alcubierre},
  \citenamefont {Nunez},\ and\ \citenamefont {Sarbach}}]{Barranco:2012qs}%
  \BibitemOpen
  \bibfield  {author} {\bibinfo {author} {\bibfnamefont {J.}~\bibnamefont
  {Barranco}}, \bibinfo {author} {\bibfnamefont {A.}~\bibnamefont {Bernal}},
  \bibinfo {author} {\bibfnamefont {J.~C.}\ \bibnamefont {Degollado}}, \bibinfo
  {author} {\bibfnamefont {A.}~\bibnamefont {Diez-Tejedor}}, \bibinfo {author}
  {\bibfnamefont {M.}~\bibnamefont {Megevand}}, \bibinfo {author}
  {\bibfnamefont {M.}~\bibnamefont {Alcubierre}}, \bibinfo {author}
  {\bibfnamefont {D.}~\bibnamefont {Nunez}}, \ and\ \bibinfo {author}
  {\bibfnamefont {O.}~\bibnamefont {Sarbach}},\ }\href {\doibase
  10.1103/PhysRevLett.109.081102} {\bibfield  {journal} {\bibinfo  {journal}
  {Phys. Rev. Lett.}\ }\textbf {\bibinfo {volume} {109}},\ \bibinfo {pages}
  {081102} (\bibinfo {year} {2012})},\ \Eprint {http://arxiv.org/abs/1207.2153}
  {arXiv:1207.2153 [gr-qc]} \BibitemShut {NoStop}%
\bibitem [{\citenamefont {Salpeter}(1964)}]{Salpeter:1964kb}%
  \BibitemOpen
  \bibfield  {author} {\bibinfo {author} {\bibfnamefont {E.~E.}\ \bibnamefont
  {Salpeter}},\ }\href {\doibase 10.1086/147973} {\bibfield  {journal}
  {\bibinfo  {journal} {Astrophys. J.}\ }\textbf {\bibinfo {volume} {140}},\
  \bibinfo {pages} {796} (\bibinfo {year} {1964})}\BibitemShut {NoStop}%
\bibitem [{\citenamefont {Bondi}\ and\ \citenamefont
  {Hoyle}(1944)}]{Bondi:1944jm}%
  \BibitemOpen
  \bibfield  {author} {\bibinfo {author} {\bibfnamefont {H.}~\bibnamefont
  {Bondi}}\ and\ \bibinfo {author} {\bibfnamefont {F.}~\bibnamefont {Hoyle}},\
  }\href@noop {} {\bibfield  {journal} {\bibinfo  {journal} {Mon. Not. Roy.
  Astron. Soc.}\ }\textbf {\bibinfo {volume} {104}},\ \bibinfo {pages} {273}
  (\bibinfo {year} {1944})}\BibitemShut {NoStop}%
\bibitem [{\citenamefont {Bondi}(1952)}]{Bondi:1952ni}%
  \BibitemOpen
  \bibfield  {author} {\bibinfo {author} {\bibfnamefont {H.}~\bibnamefont
  {Bondi}},\ }\href {\doibase 10.1093/mnras/112.2.195} {\bibfield  {journal}
  {\bibinfo  {journal} {Mon. Not. Roy. Astron. Soc.}\ }\textbf {\bibinfo
  {volume} {112}},\ \bibinfo {pages} {195} (\bibinfo {year}
  {1952})}\BibitemShut {NoStop}%
\bibitem [{\citenamefont {Edgar}(2004)}]{Edgar:2004mk}%
  \BibitemOpen
  \bibfield  {author} {\bibinfo {author} {\bibfnamefont {R.~G.}\ \bibnamefont
  {Edgar}},\ }\href {\doibase 10.1016/j.newar.2004.06.001} {\bibfield
  {journal} {\bibinfo  {journal} {New Astron. Rev.}\ }\textbf {\bibinfo
  {volume} {48}},\ \bibinfo {pages} {843} (\bibinfo {year} {2004})},\ \Eprint
  {http://arxiv.org/abs/astro-ph/0406166} {arXiv:astro-ph/0406166} \BibitemShut
  {NoStop}%
\end{thebibliography}%

\end{document}